\documentclass[12pt]{article}
\usepackage{a4wide,amsmath,amssymb,bbm}%,phystex}%,graphicx}
\usepackage{tikz}\usetikzlibrary{patterns}
\usepackage{mathrsfs,extarrows}

\newcounter{question}
\newcommand{\CC}{{\mathbb C}}
\newcommand{\NN}{{\mathbb N}}
\newcommand{\RR}{{\mathbb R}}

\newcommand{\kb}{{\boldsymbol{k}}}
\newcommand{\nb}{{\boldsymbol{n}}}

\newcommand{\xb}{\boldsymbol{x}}

\newcommand{\Mb}{{\boldsymbol{M}}}

\newcommand{\dvol}{d{\rm vol}}

\newcommand{\DD}{\mathscr{D}}
\newcommand{\FF}{\mathscr{F}}
\newcommand{\HH}{\mathscr{H}}

\newcommand{\LL}{\mathscr{L}}

\newcommand{\GG}{\mathcal{G}}

\newcommand{\Ac}{{\mathcal{A}}}
\newcommand{\Af}{{\mathcal{A}}}

\newcommand{\Zc}{{\mathcal{Z}}}

\newcommand{\CoinX}[1]{C_0^\infty({#1})}

\newcommand{\Sol}{{\sf Sol}}

\newcommand{\II}{\leavevmode\hbox{\rm{\small1\kern-3.8pt\normalsize1}}}

\newcommand{\ip}[2]{{\langle #1\mid #2\rangle}}

\newcommand{\supp}{\textrm{supp}\,}

\newcommand{\WF}{\textrm{WF}\,}
\newcommand{\Char}{\textrm{Char}\,}

\renewcommand{\Re}{\textrm{Re}\,}
\renewcommand{\Im}{\textrm{Im}\,}

\newtheorem{theorem}{Theorem}[section]
\newtheorem{definition}[theorem]{Definition}
\newtheorem{lemma}[theorem]{Lemma}

\DeclareMathOperator{\cosech}{cosech}

\newcounter{tightenum}
\newenvironment{tightitemize}% 
{\begin{list}{$\bullet$}{\setlength{\itemsep}{0pt}\setlength{\parsep}{0pt}\setlength{\topsep}{0pt}}}%
{\end{list}}
\newenvironment{tightenumerate}% 
{\begin{list}{(\roman{tightenum})}{\usecounter{tightenum} \setlength{\itemsep}{0pt}\setlength{\parsep}{0pt}\setlength{\topsep}{0pt}}}%
{\end{list}}

\begin{document}

\title{\textbf{Lectures on quantum energy inequalities}}
\author{Christopher J. Fewster\footnote{Electronic address: {\tt chris.fewster@york.ac.uk}}
\\ \textit{Department of Mathematics, University of York,}
\\ \textit{Heslington, York, YO10 5DD, UK}}
\date{\today}
\maketitle

\begin{abstract}
Quantum field theory violates all the classical energy conditions
of general relativity. Nonetheless, it turns out that quantum field theories
satisfy remnants of the classical energy conditions, known as 
Quantum Energy Inequalities (QEIs), that have been developed by various authors since the original pioneering work of Ford in 1978. 
These notes provide an introduction to QEIs and also to some
of the techniques of quantum field theory in curved spacetime
(particularly, the use of microlocal analysis together with the algebraic
formulation of QFT) that enable rigorous and general QEIs to be
derived. Specific examples are computed for the free scalar field and their consequences are discussed. QEIs are also derived for the class of unitary, positive energy conformal field theories in two spacetime dimensions. In that setting it is also possible to determine the probability
distribution for individual measurements of certain smearings
of the stress-energy tensor in the vacuum state.

\end{abstract}
%
%\begin{tikzpicture}[remember picture,overlay] \node
%[rotate=60,scale=10,text opacity=0.2] at (current page.center)
%{Draft};
%        \end{tikzpicture}
%        %\blindtext[10] 
%

\newpage
%\baselineskip=0.5\baselineskip
\tableofcontents
%\baselineskip=2\baselineskip
\newpage
\enlargethispage{\baselineskip}
\setcounter{section}{-1}
\section{Introduction and scope}

The weak energy condition (WEC) of classical general relativity 
holds if all observers measure the local energy density of matter
to be nonnegative. Expressed in terms of the  stress-energy tensor $T_{ab}$, 
the WEC amounts to the inequality $T_{ab}u^a u^b\ge 0$ for all timelike 
vectors $u^a$. This condition, and others like it, were introduced
as conditions to model generic matter distributions in the discussion
of results such as singularity theorems. The energy conditions are
obeyed by many (though not all) classical theories of interest 
and have a central place in mathematical general relativity. 

However, it has long been known that quantum field theory, our most successful fundamental matter model, is incompatible with these energy conditions~\cite{EGJ}. This raises many questions, for example: Should one doubt the validity of the singularity theorems 
for realistic matter? Can quantum fields be used to support 
`exotic' spacetime geometries (time machines, wormholes, warp drives...) which require energy-condition-violating stress-energy tensors if they are
to be solutions to the Einstein equations? 

These lecture notes provide an introduction to the subject of Quantum Energy
Inequalities (QEIs) [also often called Quantum Inequalities (QIs) in the literature]. 
These are conditions derived from within quantum field theory, that constrain
the extent to which the classical energy conditions are violated. They strongly
suggest that the answers to the questions just asked are negative 
(or raise apparently infeasible engineering problems). 

The lectures were given at the Albert Einstein Institute, Golm (March 2012)
as five lectures aimed at graduate students enrolled in the IMPRS programme. 
The secondary aim of the lectures was to provide an introduction to 
the algebraic formulation of quantum field theory in curved spacetimes and the microlocal analytic techniques that, following their
introduction by Radzikowski~\cite{Radzikowski_ulocal1996}, have been important in 
many recent developments, including the perturbative construction of
interacting field theories in curved spacetime~\cite{BrunettiFredenhagen1999, 
Ho&Wa01, Ho&Wa02, Hollands:2007zg}. From this perspective, the QEIs
provide a useful application of the theory that illustrates some of its
key features. With this in mind, the discussion of QEIs is biased towards
the derivation and analysis of various bounds, and not to 
applications such as constraints on exotic spacetime geometries. 
The reviews~\cite{Roma04,Verch_QEIreview:2008,Ford_review:2010} provide a counterpoint in that regard. 

I would like to thank the organisers of the IMPRS lectures, particularly 
Stefan Fredenhagen, for the invitation and for financial support,  the students for their interest and questions, Atsushi Higuchi and Henning Bostelmann for useful discussions and Tom Roman for comments on
the text.

\vspace{-0.3cm}
\paragraph{Summary of main conventions}
\begin{tightitemize}
\item $\hbar=c=G=1$, and the metric has signature $+--\cdots$
\item $(\nabla_a\nabla_b-\nabla_b\nabla_a)v^d = R_{abc}{}^d v^c$, 
and hence $(\nabla_a\nabla_b-\nabla_b\nabla_a)v_d = R_{abcd} v^c$;
\item the Ricci tensor is $R_{ab} = R^{d}{}_{adb}$
%\item Latin indices denote abstract indices; Greek indices denote
%coordinate components. 
\item Fourier transforms will be defined nonstandardly  by
$$
\widehat{f}(k) = \int d^nx\, e^{ik\cdot x} f(x);
$$
the hat will sometimes be displaced e.g., $f^\wedge(k)$, for typographical reasons.
\end{tightitemize}

\section{Quantum (energy) inequalities}\label{sect:QIs}

\subsection{The classical energy conditions}

The Einstein equations
\[
G_{ab} = R_{ab} - \frac{1}{2}R g_{ab} = -8\pi T_{ab} 
\]
are a curious mixture. Einstein himself likened the theory
to a building `one wing of which is built of fine marble...
but the other wing of which is built of low-grade wood'. Elsewhere,
he wrote that `the right side is a formal condensation of all things whose 
comprehension in the sense of field-theory is still problematic'.\footnote{Quoted in~\cite{capria2005physics} Ch.~ 5, p.~123.}
%\begin{quote}
%The right side is a formal condensation of all things whose 
%comprehension in the sense of field-theory is still problematic. 
%Not for a moment, of course, did I doubt that this formulation was
%merely a makeshift in order to give the general principle of relativity
%a preliminary expression. For it was essentially not anything more than a theory
%of the gravitational field, which was artificially isolated from a total field
%of as yet unknown structure.\footnote{Quoted in~\cite{capria2005physics} Ch 5, p.123.}
%\end{quote} 

Two observations illustrate some of the `low grade' features of the 
right-hand side. First, non-gravitational physics only ever considers differences between stress-energy tensors 
(or derivatives of stress-energy tensors). Arguably, the quantity on the right-hand side of the Einstein equations [the absolute stress-energy tensor] appears nowhere else in physics!\footnote{The standard 
approach--which we employ later--is to obtain the stress-energy tensor as a functional derivative of the action with respect to the metric. But many actions describe the same physics, of course.}

Second, without further qualification, the Einstein equations have no predictive power whatsoever: {\em every} smooth Lorentzian spacetime  
solves the Einstein equations for a suitable choice of $T_{ab}$ -- a fact that is often exploited in discussions of time-machines, wormholes, warp drives etc. 
A key issue is to determine what sorts of stress-energy tensors are physically reasonable. As
the real world contains a complicated combination of many different sources of stress-energy, one would ideally like some general principles that should apply fairly widely. 

The energy conditions are attempts at such principles. They are motivated partly by physics, and partly by mathematical expediency. The main conditions are:

\begin{tightitemize}
\item The Weak Energy Condition (WEC) 
\[
T_{ab} u^a u^b \ge 0 \qquad \text{for all timelike $u^a$}
\]
{\em Interpretation: All observers see nonnegative energy density.}
\item The Null Energy Condition (NEC) 
\[
T_{ab} u^a u^b \ge 0 \qquad\text{ for all null $u^a$}
\] 
\item The Dominant Energy Condition (DEC) 
\[
T_{ab} u^a v^b\ge 0 \qquad\text{for all future-pointing timelike
$u^a$ and $v^b$}
\]
{\em Interpretation: All observers see a causal flux of energy-momentum.}
\item The Strong Energy Condition (SEC) 
\[
T_{ab}u^a u^b -\frac{1}{2}g^{ab}T_{ab} \ge 0
\]
for all timelike unit $u^a$. 
\end{tightitemize}

If matter satisfies these conditions and the Einstein equations hold, we obtain 
corresponding conditions on the geometry. For example, the NEC implies that $R_{ab}u^a u^b\le 0$
for all null $u^a$, while the SEC implies the same for all timelike $u^a$. 

The principal interest in these conditions comes from the fact that they enforce various 
focussing behaviours for congruences of geodesics (see, e.g.,~\cite{HawkingEllis,Wald_gr}). Consider a congruence of timelike geodesics with future-pointing unit tangent field $u^a$. The expansion, shear and vorticity are defined uniquely
by 
\[
\nabla_b u_a = \frac{1}{3}\theta (u_a u_b-  g_{ab}) + \sigma_{ab} + \omega_{ab}
\]
and the requirements that $\sigma$ is traceless symmetric, while $\omega$ is antisymmetric. 
In particular, $\theta=\nabla_a u^a$, and $\sigma_{ab}$ is purely spatial and has
positive square $\sigma_{ab}\sigma^{ab}$. 
The Raychaudhuri equation\footnote{Derivation: Use the geodesic property of $u$ and the
Riemann tensor definition to obtain
\begin{align*}
u^c \nabla_c \nabla_b u_a &= u^c \nabla_b \nabla_c u_a -R_{cba}^{\phantom{cba}d}u^cu_d
=\nabla_b ( u^c  \nabla_c u_a ) - (\nabla_b  u^c)(\nabla_c u_a) - R_{cba}^{\phantom{cba}d}u^cu_d \\
&=- (\nabla_b  u^c)(\nabla_c u_a) - R_{cba}^{\phantom{cba}d}u^cu_d
\end{align*}
and trace to give the required result, noting that the terms in the expansion
of $\nabla_b u_a$ are mutually orthogonal.} gives 
\[
\frac{d\theta}{d\tau} =  R_{ab}u^a u^b - \frac{1}{3}\theta^2  - \sigma_{ab}\sigma^{ab} + \omega_{ab}\omega^{ab}
\]
so it is clear that an irrotational congruence obeys 
\[
\frac{d\theta}{d\tau} \le -\frac{1}{3}\theta^2
\]
if the SEC holds, or equivalently that
\[
\frac{d}{d\tau} \frac{1}{\theta} \ge \frac{1}{3}.
\]
Thus 
\[
\frac{1}{\theta(\tau)} \ge \frac{1}{\theta(0)} + \frac{\tau}{3}  
\]
and hence, 
\[
\theta(\tau) \le \frac{1}{\theta(0)^{-1} + \tau/3}.
\]
Accordingly, if $\theta(0)<0$, we see that $\theta(\tau)\to-\infty$ as $\tau\to\tau_*<3|\theta(0)|^{-1}$. 
Initially contracting irrotational congruences therefore form focal points in finite proper time
if the ambient matter distribution obeys SEC. Arguments of this type lie at the core of the singularity theorems~\cite{HawkingEllis} and many other key results in mathematical relativity. 

\paragraph{Examples} 
\begin{enumerate}
\item A perfect fluid has stress-energy tensor
\[
T_{ab} = (\mu+p) v_a v_b - p g_{ab} 
\]
for flow $4$-velocity $v^a$, energy-density $\mu$, pressure $p$. Noting that the contraction
$u^av_a$ between future-pointing timelike unit vectors $u^a$ and $v^a$ may take any value in $[1,\infty)$,  
one may easily show that 
\begin{itemize} 
\item WEC holds iff $\mu\ge 0$ and $\mu+ p\ge 0$
\item SEC holds iff $\mu+p\ge 0$ and $\mu+3p\ge 0$
\item NEC holds iff $\mu+p\ge 0$
\item DEC holds iff $\mu\ge |p|$.
\end{itemize}
{\em Exercise:} Prove these statements. Deduce that, while
it is clear that  $\text{DEC}\implies \text{WEC}\implies \text{NEC}$,
and also that $\text{SEC}\implies \text{NEC}$, there are no implications
between SEC and DEC/WEC.

\item The minimally coupled scalar field has stress-energy tensor 
\[
T_{ab}=(\nabla_a\phi) (\nabla_b \phi)
- \frac{1}{2} g_{ab}g^{cd}(\nabla_c \phi )(\nabla_d \phi) +\frac{1}{2} m^2
g_{ab}\phi^2 
\]
Given any timelike unit vector $u^a$, choose a tetrad $e_A$ with $e_0 = u$. Then 
\[
T_{ab}u^a u^b = \frac{1}{2} \sum_{A=0}^3 (e_A\cdot\nabla \phi)^2 + \frac{1}{2}m^2\phi^2 \ge 0
\]
so this theory obeys WEC due to the `sum of squares' form. Likewise, DEC also holds for this reason, 
and the same is true for the electromagnetic field. 

However, we also see that
\[
T_{ab}u^a u^b - \frac{1}{2}T^{a}_{\phantom{a}a} = (u^a\nabla_a\phi)^2 - \frac{1}{2}m^2\phi^2
\]
so the SEC can fail even for this model if $m>0$. 

\item The nonminimally coupled field, with coupling $\xi$, has stress-energy tensor
\[
T^{(\xi)}_{ab} =  T^{(\xi=0)}_{ab}  +\xi\left(g_{ab}\Box_g - \nabla_a\nabla_b - G_{ab}\right)\phi^2,
\]
As the additional terms are not of the sum of squares form, even NEC can be violated: at points where $\nabla\phi=0$, for example, this happens when
the second derivative terms outweigh contributions proportional to $\phi^2$.
\end{enumerate}

%The main aims of these lectures are:
%\begin{tightitemize}
%\item to show how QFT violates all such energy conditions...
%\item ...but nonetheless satisfies Quantum Energy Inequalities\\
%(which will be derived and analysed for free fields and conformal
%field theories in $2$-dimensions)
%\item to develop the tools of QFT in CST that are needed for
%these purposes
%\item to understand the probability distribution of individual measurements of certain smearings of the stress-energy tensor for conformal field theories.
%\end{tightitemize}
%There are many other aspects of QEIs, in terms of both
%theory and applications. I will mention some in passing, but
%otherwise will refer to the references.

\subsection{Violation of Energy Conditions in QFT}

Quantization and positivity do not mix well. For instance, the
prototypical example of a positive classical quantity --- the square
of a field --- is replaced by a Wick square in QFT. Although the
formal square is positive, it is infinite; although the Wick square is
finite, it is indefinite. 

To be specific, consider the standard quantized real scalar field
in four-dimensional Minkowski space with mass $m\ge 0$,
\[
\Phi(x) =\int \frac{d^3\kb}{(2\pi)^3\sqrt{2\omega}}
\left( e^{-ik\cdot x} a(\kb) + e^{ik\cdot x} a(\kb)^*\right),
\]
where the $4$-vector $k$ is $k_\mu=(\omega,\kb)$, with $\omega=\sqrt{\|\kb\|^2+m^2}$,
and the annihilation and creation operators obey the commutation relations
\[
[a(\kb),a(\kb')] = 0,\qquad [a(\kb),a(\kb')] = (2\pi)^3\delta^{(3)}(\kb-\kb')\II.
\]
The vacuum vector, annihilated by all the $a(\kb)$, is denoted $\Omega$. 
To form the Wick square ${:}\Phi^2{:}(x)$, of course, we modify the 
formal expression for $\Phi(x)^2$, replacing $a(\kb)a(\kb')^*$ 
by $a(\kb')^*a(\kb)$. We define the smeared Wick square by 
\[
{:}\Phi^2{:}(f) = \int d^4x\, {:}\Phi^2{:}(x) f(x),
\]
where $f\in \CoinX{\RR^4}$ is any test function on spacetime. 
Then it is a simple calculation (recalling our
convention for Fourier transform) to show that 
$$
{:}\Phi^2{:}(f)\Omega =\int \frac{d^3 \kb}{(2\pi)^3} \frac{d^3
\kb'}{(2\pi)^3} \frac{1}{2\sqrt{\omega \omega'}} 
\widehat{f}(k+k')a(\kb)^*a(\kb')^*\Omega  ;
$$
it is obvious that 
$\ip{\Omega}{{:}\Phi^2{:}(f)\Omega}=0$, and a short calculation gives
\[
\|{:}\Phi^2{:}(f)\Omega\|^2  = \int \frac{d^3 \kb}{(2\pi)^3} \frac{d^3
\kb'}{(2\pi)^3} \frac{|\widehat{f}(k+k')|^2}{2\omega \omega'} ,
\]
which is nonzero unless $f$ is identically zero\footnote{Indeed, this
is true on general grounds owing to the Reeh--Schlieder theorem.}. 
The observable ${:}\Phi^2{:}(f)$
therefore has vanishing expectation value in the state $\Omega$, but
does not annihilate $\Omega$. Standard variational arguments imply that
${:}\Phi^2{:}(f)$ must have some negative spectrum. Indeed, if we write
\[
\psi_\alpha = \cos\alpha\,\Omega + \sin\alpha {:}\Phi^2{:}(f)\Omega
\]
(assuming $f$ is chosen so $\|{:}\Phi^2{:}(f)\Omega\| = 1$) it is
easy to calculate
\[
\ip{\psi_\alpha}{{:}\Phi^2{:}(f)\psi_\alpha} = \sin(2\alpha)
+\sin^2\alpha\ip{\Omega}{{:}\Phi^2{:}(f)^3\Omega} =2\alpha + O(\alpha^2)
\]
giving negative expectation values for
sufficiently small $\alpha<0$, even if $f$ is nonnegative. 
By a scaling argument~\cite{Fews05} it
may be shown that the expectation value of ${:}\Phi^2{:}$ at a point
is unbounded from below as the state varies among Hadamard states.

The same conclusion may be reached on general grounds. An
argument due to Epstein, Glaser and Jaffe~\cite{EGJ} proves that
loss of positivity is unavoidable for Wightman fields with vanishing
vacuum expectation values. The main thrust of their argument is the following. Suppose a local 
observable $A$ has vanishing vacuum expectation value, i.e., $\ip{\Omega}{A\Omega}=0$. If $A$ is positive, it has a square root, and we have
\[
\|A^{1/2}\Omega\|^2 = \ip{\Omega}{A\Omega}=0
\]
and therefore $A^{1/2}\Omega =0$. Hence $A\Omega=0$ and, as the Reeh--Schlieder theorem~\cite{Haag} 
tells us that no nonzero local observable can annihilate the vacuum, $A$ must vanish. (The Reeh--Schlieder theorem only applies
to local observables, which is why there is no contradiction between the positivity
of the Hamiltonian and its vanishing v.e.v.) Alternatively, one can argue 
as follows: individual measurements of $A$ in state $\Omega$
constitute a random variable with vanishing expectation value; 
this implies {\em either} that zero is measured with probability $1$, in which case $A\Omega=0$ (impossible for nonzero local observables by the Reeh--Schlieder theorem) {\em or} that there is a nonzero probability for both positive and negative measurement values,
so the spectrum of $A$ extends into the negative half-line.

There are many physical situations of interest in which negative energy densities arise in QFT calculations. One of the main examples is provided by the Casimir effect, in which plane parallel conducting plates
{\em in vacuo} experience an attractive force. Actually, quite a bit can be done without much calculation
\cite{BrownMaclay:1969}. 
In the case of infinite plane plates, 
separated through distance $L$ along the $z$-axis in standard $(t,x,y,z)$ coordinates, one may
deduce on symmetry grounds and dimensional considerations that 
the stress-energy tensor of the electromagnetic field takes the form~\cite{BrownMaclay:1969}
\[
T_{\mu\nu} = \frac{C(z)}{L^4}\text{diag}\,(-1,1,1,-3).
\]
where $C(z)$ is dimensionless; here we have also used tracelessness of the stress-energy tensor. Conservation of the stress-energy tensor entails that $C(z)$ is constant except at the plates, 
so $C$ may take different values $C_0$ and $C_1$ inside and outside the plates (by reflection
symmetry the values on the two outer components should be equal). As there is no other length scale in the problem, $C_1$ and $C_0$ must be independent of the plate separation $L$. Now the two limits $L\to 0$ and $L\to\infty$ can both be regarded as describing a single plate alone in the world (as far as the outer regions are concerned). For the  stress-energy tensor outside the plates to behave in the same way in these limits, we must take $C_1=0$. The inward pressure on each plate is then $3C_0/L^4$, so we would
deduce $C_0>0$ from an attractive force. Therefore, the energy density between the plates, $-C_0/L^4$,
is negative, and we have deduced that WEC is violated. The full computation of the stress-energy tensor in QFT leads to the values $C_1=0$ and $C_0 = \pi^2/720$, replicating the Casimir force formula.\footnote{The reader might wonder how knowledge of a force, which is obtained from a difference of stress-energy tensors, has apparently allowed us to compute an absolute stress-energy tensor. The answer is that the tracelessness of the stress-energy tensor has `smuggled in' the extra information. The argument above therefore presents a choice of either accepting violations of the WEC or abandoning the standard electromagnetic stress-tensor based a conformally invariant Lagrangian.} 

One of the striking features of the Casimir result is the small magnitude of the leading coefficient $\pi^2/720 = 0.0137...$. 
Indeed, the central message of these lectures is that the energy conditions are in various ways `almost
satisfied': violations are either small in magnitude, or short-lived, or when they are not, require disparate scales, highly noninertial motion, or large positive energies somewhere in the system. Indeed, it has
been argued that the classical energy conditions might be regarded
as holding in an operational sense, once on takes account of the
positive energies present in apparatus used to produce and detect negative energy densities~\cite{Helfer_opec:1998}. 

\begin{figure}\begin{center}
\includegraphics[scale=0.4]{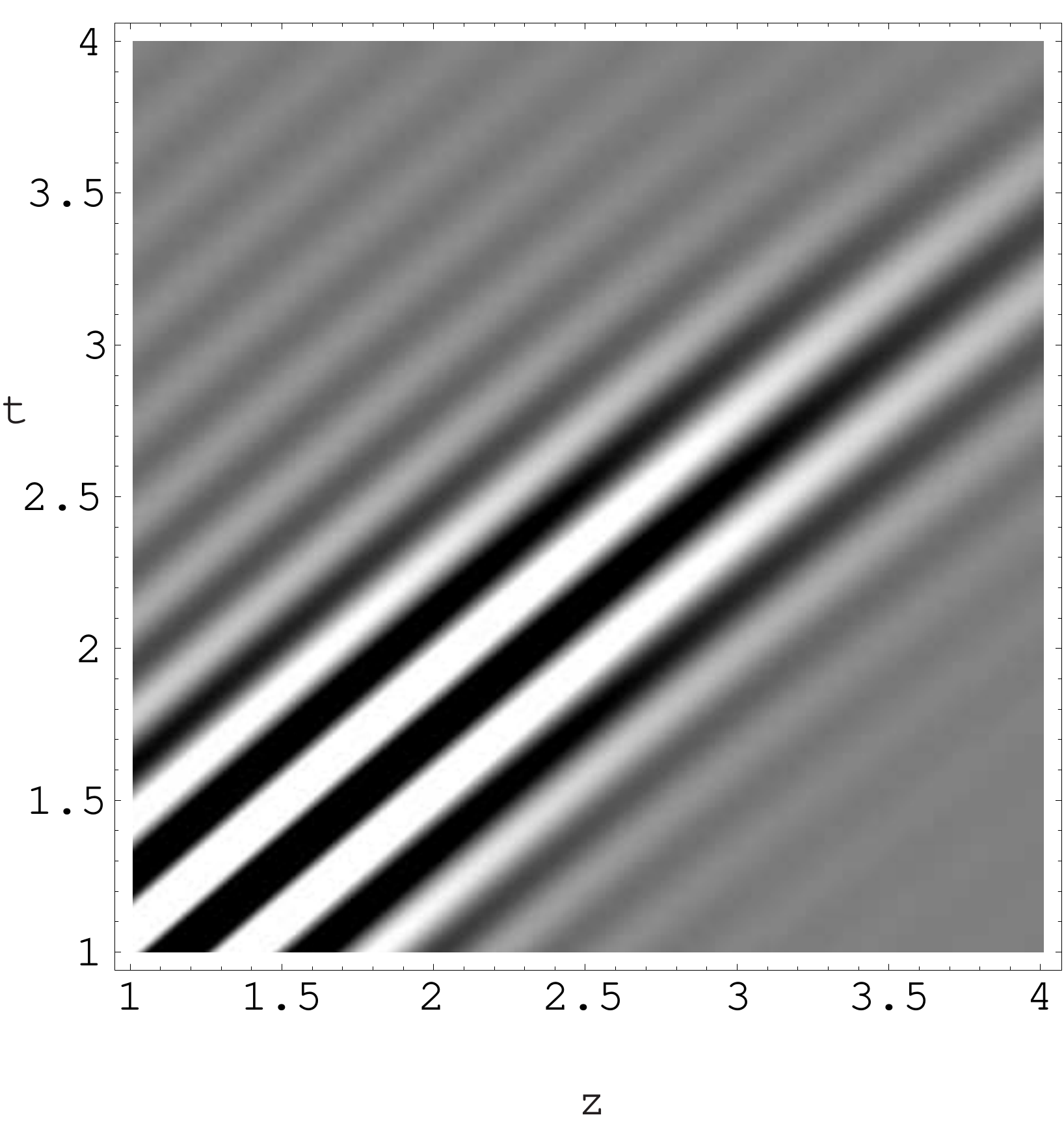}
\end{center}
\caption{A spacetime plot of the energy density in a vacuum $+$ $2$-particle superposition state~\cite{Fe&Ro03}. Dark areas represent negative values. 
\label{fig:densityplot}}
\end{figure}

Nonetheless, it is clear that there is no possibility of insisting on pointwise energy conditions
in QFT. To gain some insight into what might be possible, it is helpful to note that
the energy densities of states formed as superpositions of the vacuum and two-particle
states (like $\psi_\alpha$ above) tend to form fringes reminiscent of interference patterns. 
An example is given in Fig.~\ref{fig:densityplot} from which it can be seen that the fringes are spacelike
in character; any timelike observer meets alternating positive and negative values and 
cannot `surf' a trough of negative energy density. This suggests seeking constraints
on local averages of the energy density along timelike curves, and that is precisely what we will do.

\subsection{An example of a QEI and its consequences}\label{sect:exQEI}

The massive Klein--Gordon field in 4-dimensional Minkowski space 
obeys the following bound~\cite{FewsterEveson:1998}
\begin{equation}
\int {\langle T_{00} \rangle}_{\omega} (t, \xb) \  {|g(t)|}^{2} dt
\ge - \frac{1}{16{\pi}^{3} }\  \int_{m}^{\infty}
|\widehat{g} (u) |^{2} u^{4} 
Q_{3}(u/m) du 
\label{eq:scalarQI}
\end{equation}
for any smooth compactly supported $g$, 
and all Hadamard states\footnote{We will say more about these
states later, but for now it is enough to know that they form a large
class of physically reasonable states.} $\omega$, 
where $Q_3:[1,\infty)\to\RR^+$ is defined by
\begin{equation} 
Q_{3}(x) = \left( 1 - \frac{1}{x^2} \right)^{1/2} \left( 1 -
\frac{1}{2x^2}  \right)  -  \frac{1}{2x^4} \ln (x + \sqrt{x^{2} -
1})
\label{eq:Q3}
\end{equation}
and obeys $0\le Q_3(x)\le 1$ with
$Q_3(x)\to 1$ as $x\to\infty$. 
\begin{figure}[th]
\begin{center}
\begin{tikzpicture}%[domain=-1:0.5]
\draw[very thin, ->] (-0.5,0) -- (6,0) node[right] {$x$};
\draw[very thin, ->] (0,-0.5) -- (0,4.5) node[right] {$Q_3(x)$} ;
\draw[dashed] (0,4) -- ++(6,0);
\draw[smooth,domain=1:6,line width=1.5pt]  plot(\x,{4*(sqrt(1-(1/\x)^2)*(1-1/(2*\x^2)) - (ln(\x+sqrt(\x^2-1)))*1/(2*\x^4))});
\node[below left] at (0,0) {$0$};
\node[below] at (1,0) {$1$};
\node[left] at (0,4) {$1$};
\end{tikzpicture}
\end{center}
\caption{$Q_3(x)$}
\end{figure}

In the $m=0$ case, the bound simplifies, and actually gives
a bound valid for all $m\ge 0$ 
\begin{equation}
\int {\langle T_{00} \rangle}_{\omega} (t, \xb) \  {g(t)}^{2} dt
\ge - \frac{1}{16{\pi}^{2} }\  \int_{-\infty}^{\infty} 
|g''(t)|^{2} dt .
\label{eq:massless4dQEI}
\end{equation}
These bounds will be derived in Sec.~\ref{sec:computations} as a
special cases of more general results. 
Note that
\begin{tightitemize}
\item The left-hand side depends on the quantum state $\omega$, while the right-hand side
is state-independent.
\item The bound is known not to be optimal. 
\item The bound requires a certain degree of smoothness in $g$. In four-dimensions, it
remains valid if one take $g$ to be an element of the Sobolev space $W^{2,2}(\RR)$, 
i.e., $g$, $g'$ and $g''$ are required to exist (in the distributional sense) and be square-integrable.
But the bound does not apply to $g$ with lower regularity, in particular, to discontinuous $g$.
By `sharp switching', one can trap arbitrarily large negative energy densities. Of course, no physical device is capable of instantaneous switching, as a consequence of the uncertainty principle. 
\end{tightitemize}

The QEIs contain a lot of information, as we now show. 

\paragraph{Scaling behaviour} 
Put $g_\tau(t) = \tau^{-1/2} g(t/\tau)$. Then the bound, applied to $g_\tau$ is
\[
\frac{1}{\tau}\int {\langle T_{00} \rangle}_{\omega} (t, \xb) \  {g(t/\tau)}^{2} dt
\ge - \frac{1}{16{\pi}^{2}\tau^4}  \int_{-\infty}^{\infty}  
|g''(t)|^{2} dt ,
\]
which, in the short sampling time limit $\tau\to 0$, is 
consistent with the fact that the expectation value of energy density at a point is unbounded below, and in the limit $\tau\to\infty$ gives
\[
\liminf_{\tau\to\infty} \int {\langle T_{00} \rangle}_{\omega} (t, \xb) \  {g(t/\tau)}^{2} dt \ge 0
\]
for any Hadamard state $\omega$, 
so the WEC holds in this averaged sense (known as AWEC). 
In Sec.~\ref{sec:computations} we will see how these results are modified for noninertial 
trajectories.

\paragraph{Bounds on the duration of negative energy density}
Suppose that $\langle T_{00} \rangle_{\omega}<\rho$ for
some interval $t\in[t_0,t_0+\tau]$ of time. Then, for any $g\in\CoinX{(t_0,t_0+\tau)}$, 
\[
\rho\int |g(t)|^2\,dt \ge 
\int {\langle T_{00} \rangle}_{\omega} (t, \xb) \  {g(t)}^{2} dt
\ge - \frac{1}{16{\pi}^{2} } \int_{-\infty}^{\infty} |g''(t)|^2 dt .
\]
Rearranging and integrating by parts twice, this says that 
\[
\frac{\ip{g}{g''''}}{\ip{g}{g}} \ge -16\pi^2\rho
\]
for all such $g\neq 0$, where angle brackets denote the standard $L^2$ inner product. But the left-hand side can be minimized over $g$,
to give the minimum eigenvalue of the operator $d^4/dt^4$ on $[t_0,t_0+\tau]$,
with boundary conditions at each end corresponding to vanishing of the function
and its first derivative.\footnote{These boundary conditions emerge from
some Sobolev space analysis~\cite{FewsterTeo_Qint:2000}.} The upshot is that
\begin{equation}\label{eq:rhoCtau}
\rho \ge -\frac{C}{\tau^4},
\end{equation}
where the numerical constant $C\sim3.17$. 
Turning this around, in any time interval of duration $\tau$, the energy
density must at some instant exceed $-C/\tau^4$.  
Tighter results may be obtained for massive fields~\cite{EvFew:2007}.

\paragraph{Quantum interest}
Developing this theme, the QEI can be regarded as asserting that
the differential operator
\[
\frac{d^4}{dt^4} + 16\pi^2\rho(t) ,
\]
where $\rho(t) = {\langle T_{00} \rangle}_{\omega} (t, \xb)$, is
positive on any open interval of $\RR$, with vanishing of the function and first derivative at any boundaries. (A more precise formulation is to say that the Friedrichs extension of the above operator defined on
the dense domain $\CoinX{I}\subset L^2(I)$ is positive for any interval $I$ -- the boundary
conditions may be deduced from this~\cite{FewsterTeo_Qint:2000}.)
This leads to quite substantial restrictions
on the possible form of $\rho$. 

For example, suppose that $\rho$ has an isolated pulse, i.e., 
$\rho(t)=0$ on $[t_1-\tau_1,t_1]$ and $[t_2,t_2+\tau_2]$ with $t_1<t_2$ and $\tau_1,\tau_2>0$. 
Choose a test function $g\in\CoinX{(t_1-\tau_1,t_2+\tau_2)}$ that equals $1$ 
on $[t_1,t_2]$. Then the quantum inequality gives
\[
\int_{t_1}^{t_2} \rho(t)\,dt 
\ge -\frac{1}{16\pi^2}\left(\int_{t_1-\tau_1}^{t_1} |g''(t)|^2\,dt + 
\int_{t_2}^{t_2+\tau_2} |g''(t)|^2\,dt\right)
\]
and we can optimize over $g$ to give
\[
\int_{t_1}^{t_2} \rho(t)\,dt 
\ge -\frac{A}{16\pi^2\tau_1^3} - \frac{A}{16\pi^2\tau_2^3} \ge -\frac{A}{8\pi^2\min\{\tau_1,\tau_2\}^3},
\]
where 
\[
A = \inf_g \int_0^1 |g''(t)|^2\,dt 
\]
with $g$ restricted to smooth functions equal to $1$ near $t=0$ and $0$ near $t=1$. This amounts to an Euler--Lagrange equation $g''''=0$ with $g(0)=1$, $g'(0)=g(1)=g'(1)=0$. The solution $g(t)=1-3t^2+2t^3$ gives $A=12$, so
\[
\min\{\tau_1,\tau_2\}^3 \int_{t_1}^{t_2} \rho(t)\,dt  \ge - \frac{3}{2\pi^2},
\]
which gives a nontrivial constraint on the extent to which a pulse (of any shape) can be isolated if the integral is negative. In particular, if $\rho$ is compactly supported,
it can only be compatible with the QEI restrictions if it has nonnegative integral (another version of AWEC). Abreu and Visser~\cite{AbreuVisser:2009} have also shown that if $\rho$ is the energy density compatible with the quantum inequalities and $\int \rho = 0$, then $\rho\equiv 0$. 

Ford and Roman~\cite{FordRoman_qint:1999} first described this sort of behaviour with a financial analogy: nature allows you to `borrow' negative energy density, but you must `repay' it  
within a maximum loan term. Moreover, (excluding the case of identically zero energy density) the
amount repaid must always exceed the amount borrowed. This is the
so-called {\bf quantum interest} effect -- one may also show in various ways that
the interest rate diverges if one delays payment towards the maximum loan term.
The argument above (which is new) is based on~\cite{FewsterTeo_Qint:2000}, further developments of which can be found in~\cite{AbreuVisser:2009, TeoWong:2002}. A slightly earlier proof of some aspects
of Ford and Roman's Quantum Interest Conjecture can be found in~\cite{Pretorius:2000},
but this is not as quantitative in nature as the bounds given here.

\paragraph{Application: {\em A priori} bounds on Casimir energy densities}
Experiments conducted in a causally convex globally hyperbolic region
ought not to yield any information regarding the spacetime geometry
outside the region. This insight has been used to analyse the
Casimir effect for a long time~\cite{Kay79} and has also been at the
root of much recent progress in QFT in CST following the 
work of Brunetti, Fredenhagen and Verch~\cite{BrFrVe03}. Here, we combine
it with the quantum inequalities; what follows is based on~\cite{Few&Pfen06}
(see also~\cite{Fewster_GRG:2007}). 

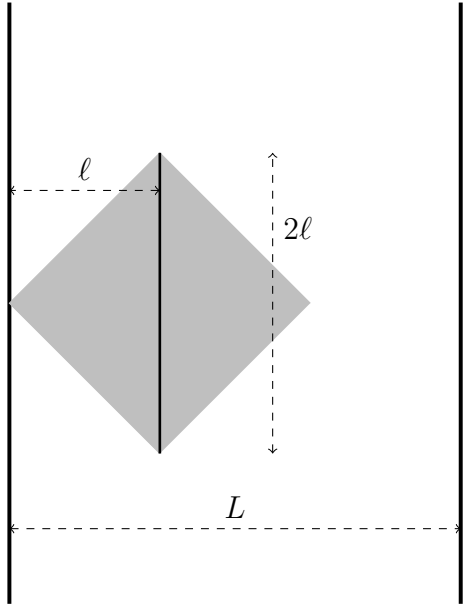
\begin{figure}[th]
\begin{center}
\begin{tikzpicture}%[domain=-1:0.5]
%\draw[dashed] (0,4) -- ++(6,0);
\draw[line width =1.5pt] (-3,-4) -- (-3,4);
\draw[line width =1.5pt] (3,-4) -- (3,4);
\draw[fill=lightgray, lightgray] (-1,-2) -- ++ (-2,2) -- ++ (2,2) -- ++(2,-2) ++(-2,-2); 
\draw[line width =1pt] (-1,-2) -- ++(0,4);
\draw[<->, dashed] (-1,1.5) -- node[above]{$\ell$} ++(-2,0);
\draw[<->, dashed] (-3,-3) -- node[above]{$L$} ++(6,0);
\draw[<->, dashed] (.5,-2) -- node[right, near end]{$2\ell$} ++ (0,4);
%\draw[smooth,domain=-4:4,line width=1.5pt]  plot(\x,{sqrt(\x^2+4)});
%\draw[dashed] (-4,4) -- (0,0) -- (4,4);
%\def\mypath{(-4,1) -- (0,0) -- (4,1) -- (4,-2) -- (-4,-2) -- (-4,1)};
%\fill[pattern = crosshatch] \mypath;
%\draw[very thin, ->] (0,-2) -- (0,5) node[right] {$k_0$};
%\draw[very thin, ->] (-4,0) -- (4.2,0) node[right] {$\kb$};
%\node[below left] at (0,0) {$0$};
%\node[below] at (1,0) {$1$};
%\node[left] at (0,4) {$1$};
\end{tikzpicture}
\end{center}
\caption{Spacetime diagram of the Casimir plate set-up.}
\end{figure}

Consider a region between Casimir
plates  at $z=\pm L/2$ in otherwise flat spacetime and an inertial trajectory parallel to the plates. Let $\ell$ be the distance from this trajectory to the nearest plate in the $t=0$ surface. Then no experiment conducted along the
trajectory in a time interval of less than $2\ell$ can possibly know about the
existence of the plates; it should be as if the experiment was conducted in Minkowski space. In particular the quantum energy inequalities apply, and [as the energy density is supposed constant along the trajectory] 
\eqref{eq:rhoCtau} gives an {\em a priori} bound
\[
T_{00} \ge -\frac{C}{(2\ell)^4} \sim -\frac{3.17}{(2\ell)^4} = -\frac{0.20}{\ell^4}. 
\]
By comparison, the known value of the Casimir energy density
for the massless minimally coupled scalar field is 
\[
T_{00} = -\frac{\pi^2}{1140L^4} -
\frac{\pi^2}{48L^4}\,\frac{3-2\cos^2(\pi z/L)}{\cos^4 \pi z/L}
\]
which ranges between $3$-$7\%$ of the bound as $z$ varies in $[-L/2,L/2]$. 

However, the a priori bound is valid even in situations where exact
calculation is difficult/impossible; it also applies to all stationary (Hadamard) states
of the system. This partly answers the question (often emphasised by Ford): 
why are the Casimir energies so small? They are constrained by QEIs,
which already gives a small leading constant in front of the $1/\ell^4$ 
one might expect on dimensional grounds. There remains an interesting
question as to why the Casimir energy density is a comparatively small
proportion of the allowed bound.

\subsection{Some history and references}

The study of QEIs began with a 1978 paper of Ford~\cite{Ford78}, in which he argued
that a beam of negative energy (described by a pure quantum state) could be used to cool a hot body and decrease its entropy. Ford argued from the
macroscopic validity of the second law of thermodynamics that violations of the energy conditions must be governed by bounds of
uncertainty principle type. That was borne out in subsequent derivations
of quantum inequalities by Ford in a series of papers written in conjunction with Roman and Pfenning~\cite{Ford:1991,FordRoman:1995,FordRoman:1997,PfenningFord_RW:1997, PfenningFord_static:1998,FordPfenningRoman:1998}, concerning Minkowski space and some 
static spacetimes. These papers established lower bounds on 
weighted averages of the energy density\footnote{In fact, the
earliest papers consider negative energy-momentum fluxes, but the
energy density soon became the main object of interest.} of the scalar and electromagnetic quantum fields along a static trajectory, where the weight is given by the Lorentzian function
$f(t) = \tau/(\pi(t^2+\tau^2))$. Here $\tau$ sets the timescale
for the averaging. For example, the massless scalar field in $4$-dimensions obeys a bound
\[
\int \frac{\tau\langle T_{00}(t,\xb)\rangle_\omega}{\pi(t^2+\tau^2)}dt 
\ge -\frac{3}{32\pi^2\tau^4}
\]
for all sufficiently nice states $\omega$ and any $\tau>0$. 

The first QEI for general weighted averages was derived by 
Flanagan~\cite{Flanagan:1997} for the special case of
massless quantum fields in two-dimensional
Minkowski space. His argument forms the basis of a
general argument for two-dimensional conformal field theories
\cite{Fe&Ho05} that will be discussed in Sect.~\ref{sect:CFT_QEI}.

The bound discussed in Sect.~\ref{sect:exQEI} was derived in~\cite{FewsterEveson:1998} 
for the scalar field of mass $m\ge 0$ in Minkowski space of arbitrary dimension and for averaging along inertial curves with general weight functions of sufficiently rapid decay. This
was generalized to some static spacetimes~\cite{FewsterTeo:1999}
for averaging along static trajectories. With some modification,
the method also applies to the electromagnetic~\cite{Pfenning:2001}, Dirac~\cite{FewsterMistry:2003}
and Rarita--Schwinger fields~\cite{YuWu:2004}. The general approach
of~\cite{FewsterEveson:1998} (somewhat rephrased) 
formed the basis for the first fully rigorous QEI~\cite{Fews00}
for the scalar field, which was also much more general than the previously known results. We will discuss that argument in 
Sect.~\ref{sect:QEI_deriv}.  Generalizations to the Dirac
and electromagnetic fields are also known~\cite{FewsterVerch_Dirac,DawsFews06,Few&Pfen03}.

There is a significant literature on the theory and applications of QEIs---reviews can be found in  \cite{Fews05,Few:Bros,Roma04} and the recently published~\cite{EvRom_book} gives a popular but nonetheless careful account.
I mention only two aspects here. First, QEIs place significant constraints on the ability of quantum fields to support wormholes or other exotic spacetimes, if the fields are assumed to obey a QEI similar to those found for the free scalar fields~\cite{PfenningFord_warp,FordRoman_worm,FewsterRoman_worm}. Second, the
link between QEIs and thermodynamics, which originally motivated
Ford~\cite{Ford78}, can be pursued abstractly (in a setting that
includes the scalar field)~\cite{FewsterVerch_passive}.

\section{Some methods of Quantum Field Theory in Curved spacetime}

The QEI studied in Sec.~\ref{sect:exQEI} can be derived directly by fairly elementary
means~\cite{FewsterEveson:1998}. However, it is also a special case of a rather general
QEI, whose proof will be our main goal. To achieve this we introduce
the algebraic formulation of QFT in CST; the completion of the proof will also need the tools of microlocal analysis in a subsequent lecture. In terms of literature, \cite{BarFredenhagen_LNP} contains
much relevant material, while \cite{Wald_qft} emphasises a slightly different version of the algebraic approach (and does not cover microlocal analytic methods). 
Some of this material is based on~\cite{Fewster_Leipzignotes:2008} (although emphases differ) which contains
a more broadly based account of QFT in CST.\footnote{I am aware of
a number of misprints and minor errors in~\cite{Fewster_Leipzignotes:2008}, which I hope to correct in due course.}

Throughout, let $\Mb$ be a {\bf globally hyperbolic
spacetime}, understood to comprise a (smooth etc) $n$-dimensional manifold
with time-oriented Lorentz metric and such that
\begin{tightitemize}
\item there are no closed causal curves
\item for any points $p,q$, the intersection of causal futures/pasts
$J^+(p)\cap J^-(q)$ is compact.\footnote{The definition
given in~\cite{HawkingEllis,Wald_gr} also requires strong causality, but this is
a consequence of the other two conditions and can be dropped~\cite[Thm~3.2]{Bernal:2006xf}.}
\end{tightitemize}

\subsection{The Klein--Gordon field}

We will study the formulation of the real scalar field, defined by Lagrangian density
$$
\LL_g[\phi] =\frac{1}{2}\rho_g \left( g^{ab} (\nabla_a\phi)(\nabla_b \phi)
-(m^2+\xi R)\phi^2\right),
$$
where $\rho_g$ is the density induced by the metric $g_{ab}$ of $\Mb$, $R$ is the Ricci scalar and
$\xi$ is a dimensionless coupling constant. The case $\xi=0$ is known as {\bf minimal coupling} and
$\xi\not=0$ as {\bf non-minimal coupling}.  In the special case $m=0$, $\xi=(n-2)/(4n-4)$, the action exhibits conformal invariance, because the Lagrangian density is unchanged under the simultaneous replacements
$$
g_{ab}\to \overline{g}_{ab} = \Omega^2 g_{ab} \qquad \phi\to \overline{\phi}=\Omega^{1-n/2}\phi
$$
for any smooth positive function $\Omega$, i.e.,
$\LL_{\overline{g}}[\overline{\phi}]=\LL_g[\phi]$. This value of $\xi$ is accordingly called
{\bf conformal coupling}.

This field equation derived from this action is the Klein--Gordon equation 
$$
P\phi:=(\Box_g +m^2 + \xi R)\phi = 0, \qquad \text{where}~\Box_g = g^{ab}\nabla_a\nabla_b,
$$
and the stress-energy tensor, obtained by varying the action with respect to the metric, is
\begin{align*}
T_{ab}&=(\nabla_a\varphi) (\nabla_b \varphi)
- \frac{1}{2} g_{ab}g^{cd}(\nabla_c \varphi )(\nabla_d \varphi) +\frac{1}{2} m^2
g_{ab}\varphi^2 \\
&\quad +\xi\left(g_{ab}\Box_g - \nabla_a\nabla_b - G_{ab}\right)\phi^2,
\end{align*}
where $G_{ab}$ is the Einstein tensor. Note that the effect of the
coupling constant can be seen in the stress-energy tensor even where
the metric is Ricci flat, even though the $\xi R\phi^2$ term in the Klein--Gordon
equation vanishes in such situations.

The Klein--Gordon field is well-posed on globally hyperbolic spacetimes,
for which we refer to the thorough and clear presentation of~\cite{BarGinouxPfaffle}.
For our purposes, the main result is: 
\begin{theorem} If $\Mb$ is globally hyperbolic then, to each $f\in\CoinX{\Mb}$ there exists
$\phi^\pm\in C^\infty(\Mb)$, with $\supp \phi^\pm\subset J^\pm(\supp f)$, solving the inhomogeneous problem
\begin{equation}\label{eq:KG_inhom}
P\phi^\pm = f,
\end{equation}
Moreover, $\phi^{+/-}$ is the unique (distributional) solution to \eqref{eq:KG_inhom} whose support
is past/future-compact (i.e., the support has compact intersection with every set of the form $J^\mp(p)$). 
The maps 
\begin{align*}
E^\pm: \CoinX{\Mb} & \longrightarrow C^\infty(\Mb)  \\
f &\longmapsto \phi^\pm
\end{align*}
are linear continuous mappings, where $\CoinX{\Mb}$ and $C^\infty(\Mb)$ are given
their standard topologies.
\end{theorem}
Due to the support properties, $E^-$ (resp., $E^+$) is called the {\bf
advanced} (resp., {\bf retarded}) {\bf fundamental solution} (or Green
function). In the special case where $f=Pf'$ for some $f'\in\CoinX{\Mb}$,
we note that $f'$ is both past and future compact, so $f'=E^\pm f$ by uniqueness. Hence we have
\[
E^\pm P f' = f'
\]
together with the initial property $P E^\pm f = f$.

The {\bf advanced-minus-retarded fundamental solution} $E$ is defined by
$E=E^--E^+$. (Warning: some authors use retarded-minus-advanced, or
label retarded and advanced the other way round! Furthermore, in the
$-+++$ signature, the fundamental solutions to $(\Box_g-m^2)\phi=f$ are
\emph{minus} the fundamental solutions we use; e.g., Wald's $A$~\cite{Wald_qft} is our
$-E^-$.)  Clearly $\phi = Ef$ is a
smooth solution to the homogeneous equation $P\phi=0$, but we also have
an important result (cf.~\cite[Thm 3.4.7]{BarGinouxPfaffle}) that summarises
a number of key properties in a compact form.
\begin{theorem} \label{thm:exact}
The following is an exact sequence (that is, the image of each map is precisely equal to the kernel of the next):
\begin{equation}\label{eq:exact}
0\longrightarrow \CoinX{\Mb} \xlongrightarrow{P} \CoinX{\Mb}  \xlongrightarrow{E} C^\infty_{SC}(\Mb) 
\xlongrightarrow{P} C^\infty_{SC}(\Mb)
\end{equation}
where $C^\infty_{SC}(\Mb)$ denotes those functions in $C^\infty(\Mb)$ with support contained in 
$J(K)=J^+(K)\cup J^-(K)$ for some compact $K$.
\end{theorem}
{\bf Remark:} The support of any function $F\in C^\infty_{SC}(\Mb)$ has compact intersection with 
any Cauchy surface. But it is not the case that a smooth function whose support has compact intersection with each leaf of a given foliation of $\Mb$ by Cauchy surfaces is necessarily in $C^\infty_{SC}(\Mb)$.\footnote{For example, in four dimensional Minkowski space, 
the set $\cup_{n=1}^\infty \{1/n\}\times B_{n}$, where  $B_n$ is a ball
of unit radius, centred at $(4n,0,0)\in\RR^3$ has compact intersection with each $t=\text{const}$ 
hypersurface, but is not contained in $J(K)$ for any compact $K$.} Unfortunately the literature contains many references to functions `compactly supported on Cauchy surfaces' that would be more accurately rendered as `in $C^\infty_{SC}(\Mb)$'.\\
{\em Proof:} 
The equalities  $EPf=E^-Pf - E^+Pf = f-f =0$ and $PEf = PE^-f - PE^+f = f - f =0$ for $f\in\CoinX{\Mb}$ are immediate, so each image is certainly contained in the kernel of the following map. For the reverse inclusions, we observe that 
\begin{itemize}
\item if $Pf=0$ with $f\in\CoinX{\Mb}$ then $f=E^+0=0$ by uniqueness of past-compact solutions; 
\item if $Ef=0$ with $f\in\CoinX{\Mb}$ then
$E^+f = E^-f$, which shows that $E^+f$ is supported in the compact set
$J^+(\supp f)\cap J^-(\supp f)$ and hence $f=PE^+f\in P \CoinX{\Mb}$; 
\item if $P\phi=0$ with $\phi\in C^\infty_{SC}(\Mb)$ we argue as follows. Choose any two 
Cauchy surfaces $\Sigma^\pm$, with $\Sigma^+\subset I^+(\Sigma^-)$ and a smooth function
$\chi$ with $\chi=1$ in $J^-(\Sigma^-)$ and $\chi=0$ in $J^+(\Sigma^+)$. Then
\[
f = P\chi \phi
\]
is compactly supported (in $I^+(\Sigma^-)\cap I^-(\Sigma^+)$). As $\chi\phi$ has
future-compact support, $\chi\phi = E^- f$. But $(\chi-1)\phi$ has past-compact support, and
$P(\chi-1)\phi= f$, so $(\chi-1)\phi= E^+f$. Subtracting, $\phi=E^-f - E^+f = Ef$. \\
{\em Note:} This shows that if $O$ is any open neighbourhood of a Cauchy surface, then
any solution $\phi$ may be expressed as $Ef$ for some $f\in\CoinX{O}$. 
\end{itemize}
$\square$

\subsection{Phase space}

\paragraph{The symplectic space}
Our phase space consists of all real-valued solutions with $SC$ support
\[
\Sol_\RR(\Mb) := \{\phi\in C^\infty_{SC}(\Mb;\RR): P\phi =0\}.
\]
However, it is rather convenient to work with its complexification, i.e., the space of 
complex-valued solutions
\[
\Sol(\Mb) := \{\phi\in C^\infty_{SC}(\Mb): P\phi =0\}.
\]
In view of the exact sequence~\eqref{eq:exact} and the first isomorphism theorem for vector spaces, 
this may be reformulated as
\[
\Sol(\Mb) = E\CoinX{\Mb} \cong \CoinX{\Mb}/P\CoinX{\Mb}.
\]
Let us write $\hat{E}$ for the isomorphism $\hat{E}:\CoinX{\Mb}/P\CoinX{\Mb}\to \Sol(\Mb)$,
which has action
\[
\hat{E} (f + P\CoinX{\Mb}) = Ef .
\]

The pairing $C^\infty(\Mb)\times \CoinX{\Mb}\to \CC$,
\[
(\phi,f)\longmapsto \int_\Mb \phi f\dvol
\]
clearly induces a pairing $\Sol(\Mb)\times (\CoinX{\Mb}/P\CoinX{\Mb})\to \CC$ by
\[
\langle \phi,[f]\rangle = \int_\Mb \phi f \dvol
\]
(we use the formal self-adjointness of $P$ here). We may now define a bilinear map on $\Sol(\Mb)$ by
\[
\sigma(\phi_1,\phi_2) = \langle\phi_2, \hat{E}^{-1}\phi_1\rangle , \qquad \text{(note reversal of order!)}
\]
which evidently has the properties that
\begin{equation}\label{eq:sigma_prop}
\sigma(E f,\phi)=  \langle \phi , \hat{E}^{-1}Ef\rangle = \langle\phi,[f]\rangle = \int_\Mb \phi f\dvol
\end{equation}
and hence
\[
\sigma(E f_1, Ef_2) = \int_\Mb (Ef_2)f_1 \dvol \stackrel{\text{def}}{=}E(f_1,f_2).
\]

An easy calculation shows that
\begin{equation}\label{eq:sig_formula}
\sigma(\phi_1,\phi_2) =\int_\Sigma \left(\phi_1 \nabla_\nb \phi_2 - \phi_2 \nabla_\nb\phi_1\right) d\Sigma
\end{equation}
for any smooth spacelike Cauchy surface $\Sigma$ with future-pointing unit normal vector $\nb$, from 
which it is clear that $\sigma$ is antisymmetric and, moreover, is the standard symplectic form for
the Klein--Gordon system. To prove~\eqref{eq:sig_formula}, write $\phi_1 = Ef$ for some $f\in\CoinX{M}$ supported to the past of $\Sigma$. Given the definition of $E$ and the support properties of $E^\pm$, $\phi_1 = (E^- -E^+)f = -E^+f$ on $\Sigma$, so
\begin{align*}
\text{RHS of~\eqref{eq:sig_formula}} &=\int_\Sigma \left( - E^+f \nabla_\nb\phi_2
+\phi_2 \nabla_\nb E^+f \right) d\Sigma \\ 
& = \int_{I^-(\Sigma)} \nabla^a \left(\phi_2 \nabla_a E^+f - E^+f \nabla_a \phi_2\right)\dvol  \\  
&= \int_{I^-(\Sigma)} \left(\phi_2 P E^+f - (E^+f) P\phi_2\right) \dvol\\
&= \int_\Mb \phi_2 f \dvol= \sigma(\phi_1,\phi_2)
\end{align*}
using the divergence theorem in conjunction with the fact that $E^+f$ has past-compact support, 
and $\supp f\subset I^-(\Sigma)$, together with Eq.~\eqref{eq:sigma_prop}.

The map $\sigma$ is evidently {\bf weakly nondegenerate}, in the sense that if $\sigma(\phi',\phi) =0$ for all $\phi'\in\Sol(\Mb)$, then, putting $\phi'=Ef$, we find $\int \phi f \dvol = 0$ for all $f\in\CoinX{\Mb}$ and hence $\phi=0$. 

As mentioned, $\sigma$ is the standard symplectic form for the Klein--Gordon field. To make 
our conventions more explicit, we observe that the covariant momentum conjugate $\Pi^a$ to the field is
defined by the functional derivative\footnote{The meaning of this expression is that
\[
\int_\Mb \Pi^a w_a\,\dvol =\left.\frac{d}{d\lambda}\int_V \LL(\phi,\nabla\phi+\lambda w)\right|_{\lambda=0}
\]
for every smooth compactly supported covector field $w_a$, and any relatively compact open subset $V$
containing $\supp w$.}
\[
\Pi^a = \frac{1}{\rho_g} \frac{\delta S}{\delta \nabla_a\phi} =
 \nabla^a\phi
\]
so
\[
\sigma(\phi_1,\phi_2) =\int_\Sigma \left(\phi_1  \Pi_2^a  - \phi_2  \Pi_1^a\right)n_a d\Sigma.
\]
(This corresponds to the convention that the symplectic form, in finite dimensions, may be written $\sigma= dq^i\wedge dp_i$ in terms of canonical coordinates $q^i$ and momenta $p_i$.)

The upshot is that $\Sol_\RR(\Mb)$, equipped with (the restriction of) $\sigma$ is a weakly nondegenerate symplectic space, while $\Sol(\Mb)$ equipped with $\sigma$ and complex conjugation, is its complexification. 

\paragraph{Classical observables and Poisson brackets}

Classical observables are functions on this phase space: for example, every $f\in\CoinX{M;\RR}$
defines an observable $F_f$ which acts on solutions 
$\phi\in\Sol_\RR(\Mb)$ by
$$
F_f(\phi) = \int_\Mb\dvol_g(p) \phi(p)f(p) = \sigma(Ef,\phi).
$$
(By weak nondegeneracy, the last equality shows that there are enough
such observables to distinguish elements of $\Sol_\RR(M)$.) 
We may observe that the $F_f$ depend linearly on $f$, and that
some of them vanish identically:
\[
F_{Pf}(\phi) = \sigma(EPf,\phi) = 0
\]
for any $f\in\CoinX{M;\RR}$ and $\phi\in\Sol_\RR(M)$. 

If $(\mathscr{P},\omega)$ is a finite-dimensional (real) symplectic manifold, the Poisson bracket of two smooth functions
$F,G\in C^\infty(\mathscr{P})$ is given in terms of the exterior derivatives of $F$ and $G$ by
\begin{equation}
\{F,G\}(p)=dF(dG^\sharp)|_{p},
\end{equation}
where $(dG)^\sharp|_p\in T_p\mathscr{P}$, which is the Hamiltonian vector
field induced by $G$, satisfies
\begin{equation}
\omega_{p}((dG)^\sharp|_{p},v)=dG|_{p}(v)
\end{equation}
for $v \in T_{p}\mathscr{P}(M)$ according to our convention for the symplectic form.\footnote{
Our convention for Poisson brackets then amounts to
\[
\{F,G\} = \frac{\partial F}{\partial q^i}\frac{\partial G}{\partial p_i} - \frac{\partial F}{\partial p_i}\frac{\partial G}{\partial q^i}
\]
in canonical coordinates in the finite dimensional case.}
In particular, if $\mathscr{P}$ is a vector
space (regarded as a manifold with $T_p\mathscr{P}\cong \mathscr{P}$) 
and $F$ and $G$ are linear functionals on $\mathscr{P}$, then
$dF|_p(v) = F(v)$, etc, so the Poisson bracket--a function on phase space--is a constant,
\[
\{F,G\}\equiv F(G^\sharp), \qquad \text{where}\quad \omega(G^\sharp,v)=G(v).
\]
Although infinite-dimensional manifolds require care, these formulae will be enough for our purposes.  With $G=F_f$, we know that 
\[
F_f(\phi) =\sigma(Ef,\phi)
\]
so we may take $F_f^\sharp =Ef$ (and there is no other solution, by weak nondegeneracy). Hence
\[
\{F_{f_1}, F_{f_2}\} \equiv F_{f_1}(Ef_2) = \int_\Mb f_1 Ef_2 \dvol= E(f_1,f_2).
\]

It is worth observing that our class of observables $\FF(\Mb):=\{F_f:f\in \CoinX{\Mb;\RR}\}$ is, itself, a copy of the phase space, when equipped with the Poisson bracket as the symplectic form: the map $F_f\mapsto Ef$ is easily seen to be a symplectic isomorphism. So this class of observables provides a complete description of the  underlying dynamical system.

\subsection{Algebraic formulation of the quantum field theory}

The algebraic approach is actually nothing but Dirac quantization, but without
requiring quantized observables to act on a Hilbert space in the first instance.

\paragraph{Dirac quantization}
Applying Dirac's quantization prescription to the classical
observables $F_f$, we seek (at least formally) self-adjoint operators\footnote{We will write
hats on top of operators only in this section. This should not be confused with the notation
for a Fourier transform used later.} $\widehat{F_f}$
($f\in\CoinX{M;\RR}$) obeying the same algebraic relations as the $F_f$, but with the
standard replacement of Poisson brackets by commutators
\begin{equation}\label{eq:commutators}
[\widehat{F_f},\widehat{F_{f'}}] =i\widehat{\{F_f,F_{f'}\}} = iE(f,f')\II 
\end{equation}
(note that any constant function is quantized as an appropriate multiple of the unit $\II$).
As quantizations of the classical smeared fields, the $\widehat{F_f}$ are interpreted as smeared quantum fields.
In particular, when the supports of $f$ and $f'$ are
spacelike-separated, $\widehat{F_f}$ and $\widehat{F_{f'}}$ should 
commute, reflecting the Bose statistics of a spin-$0$ field. 

It is also convenient to permit smearings with complex-valued functions.
Accordingly, we define
$$
\Phi(f) = \widehat{F_{\Re f}} + i\widehat{F_{\Im f}}
$$
for $f\in\CoinX{M}$, dropping the hats from now on
and seek to implement the following relations):
\begin{itemize}
\item $f\mapsto \Phi(f)$ is complex-linear;
\item $\Phi(f)^*=\Phi(\overline{f})$ for all $f\in\CoinX{M}$
\item $\Phi(Pf) =0$ for all $f\in\CoinX{M;\RR}$ for all $f\in\CoinX{M}$
\item $[\Phi(f),\Phi(f')]=iE(f,f')\II$ for all $f,f'\in\CoinX{M}$.
\end{itemize}
This may be done by invoking a unital $*$-algebra with abstract elements
$\Phi(f)$ ($f\in\CoinX{\Mb})$ as generators, subject to the above relations. 
We denote it $\Af(\Mb)$. (The only risk is that $\Af(\Mb)$ might be trivial, but it is 
not: as a vector space it is isomorphic to the symmetric tensor vector space
\[
\Gamma_\odot (\Sol(\Mb)) = \bigoplus_{n=0}^\infty \Sol(\Mb)^{\odot n}
\]
over the solution space $\Sol(\Mb)$ on $\Mb$. In fact, $\Af(\Mb)$ is also
simple, so one could not impose additional relations without it collapsing to 
the trivial algebra.)

\paragraph{States and the GNS representation}

Self-adjoint elements [$A^*=A$] of $\Ac(\Mb)$ should play the role of
observables. However, this is rather empty without a rule for turning observables into
expectation values, in other words,  notion of a state. 

\begin{definition} A state on $\Ac(\Mb)$ is a linear map
$\omega:\Ac(\Mb)\to\CC$ obeying
\begin{align*}
\omega(\II)=1 & \qquad\textrm{normalisation}\\
\forall A\in\Ac(\Mb),~\omega(A^*A)\ge 0 & \qquad \textrm{positivity}.
\end{align*}
\end{definition}

Expectation values 
\[
\omega_n(f_1,\ldots,f_n) \stackrel{\rm def}{=}\omega(\Phi(f_1)\Phi(f_2)\cdots\Phi(f_n))
\]
are called $n$-point functions. It is clearly sufficient to specify the $n$-point
functions to fix $\omega$. The algebraic relations in $\Mb$ have implications for the $n$-point functions: for example, 
\[
\omega_2(f_1,f_2) -  \omega_2(f_2,f_1)= \omega(\Phi(f_1)\Phi(f_2)-\Phi(f_2)\Phi(f_1)) =
\omega(iE(f_1,f_2)\II) = iE(f_1,f_2),
\]
and
\[
\omega_2(Pf_1,f_2) = \omega(\Phi(Pf_1)\Phi(f_2)) = 0 = \omega_2(f_1,Pf_2)
\]
while positivity of the state implies directly that
\[
\omega_2(\overline{f},f) = \omega(\Phi(\overline{f})\Phi(f)) = 
\omega(\Phi(f)^*\Phi(f) \ge 0. 
\]
Thus $\omega_2(x,x')$ is a  bidistribution of {\bf positive type} that is a bisolution 
to the Klein--Gordon equation and whose antisymmetric
part is $\frac{i}{2}E(x,x')$.

Perhaps reassuringly, given a state $\omega$ we may regain a Hilbert
space setting using the {\bf GNS construction} (Gel'fand, Naimark, Segal)
which gives a Hilbert space $\HH_\omega$, a dense domain $\DD_\omega\subset\HH_\omega$, a representation
$\pi_\omega$ of $\Ac(\Mb)$ as (generally unbounded) operators defined on $\DD_\omega$, and a 
distinguished vector $\Omega_\omega\in\DD_\omega$ such that
\[
\ip{\Omega_\omega}{\pi_\omega(A)\Omega_\omega} = \omega(A) 
\]
for all $A\in \Ac(\Mb)$. However, we will not develop this here; see, e.g.~\cite{Haag}. 

\paragraph{Hadamard states}

The algebra $\Af(\Mb)$ admits rather too many states and it is
necessary to select a `physically reasonable' subclass. We consider
{\bf Hadamard states} which are states whose $2$-point functions
are distributions and take a specific form for near-coincidence 
of the points. The Hadamard class was precisely described in~\cite{KayWald1991};
the definition given there is rather involved, but the rough idea (in four spacetime dimensions) 
is that when $x$ and $x'$ lie in a common causally convex geodesic normal neighbourhood,
one should have
\begin{equation}\label{eq:Hadamard}
\omega_2(x,x') = 
\frac{U(x,x')}{4\pi^2\sigma_+(x,x')} + V(x,x')\log(\sigma_+/\ell^2) +W(x,x')
\end{equation}
where $U$, $V$ and $W$ are smooth, $\sigma$ the signed square geodesic separation of $x$
and $x'$, taken to be positive for spacelike separation\footnote{Note: Some authors, including~\cite{dWB:1960}, use
$\sigma$ for {\em half} of the signed squared separation; our convention follows e.g.,~\cite{KayWald1991}.} and the notation
$f(\sigma_+)$ indicates a certain regularization of
$f(\sigma)$ (the Minkowski space case is given explicitly below). 
The parameter $\ell$ is a length scale, necessary for dimensional reasons.  
The functions $U$ and $V$ are defined using $\ell$, the local geometry
and the Klein--Gordon operator, along with the condition that
$U(x,x)=1$, which allow $U$ to be identified as the 
square root of the van Vleck-Morette determinant,\footnote{See~\cite[\S 1]{dWB:1960},
modulo change in notation. Here the derivatives are {\em partial derivatives} in some coordinate system,
and $g=|\det g_{\mu\nu}|$ in the same coordinates. {\em Exercise:} check
that this is a bi-scalar quantity.}
\[
\Delta(x,x') = -
\frac{\det [-\partial_\mu \partial_{\mu'}\frac{1}{2}\sigma(x,x')]}{\sqrt{g(x)g(x')}}
\]
and $V$ to be expressed as series in $\sigma$. 
(In general, the series for $V$ does not converge but there are various standard work-arounds that 
I will not discuss here.) 
All the state-dependent information is contained in $W$. There is a much cleaner definition of the Hadamard class
in terms of microlocal analysis -- see Sec.~\ref{sect:uSC}.

The motivation for Eq.~\eqref{eq:Hadamard} is that it makes the singular part of $\omega_2$ as
much as possible like the leading behaviour of the Minkowski vacuum $2$-point function,
which, for mass $m\ge 0$ is 
\[
\omega_2(x,x') =  \lim_{\epsilon\to 0^+}
F([-(t-t'-i\epsilon)^2+|\xb-\xb'|^2]^{1/2}),
\]
where, for $m>0$,
\[
F(z) = \frac{m}{4\pi^2 z} K_1(mz) = \frac{1}{4\pi^2 z^2} +\frac{m^2}{8\pi^2}\log(mz)+ O(1)
\]
while $F(z)=1/(4\pi^2 z^2)$ for $m=0$.
Hence 
\[
\omega_2(x,x')  = \frac{1}{4\pi^2\sigma_+}
+\frac{m^2}{16\pi^2}\log m^2\sigma_+ +\ldots,
\]
where $f(\sigma_+)$ is the distributional limit
\[
f(\sigma_+) = \lim_{\epsilon\to 0^+} f(-(t-t'-i\epsilon)^2+|\xb-\xb'|^2).
\]
The major consequence of the definition is that the difference of two Hadamard $2$-point functions
is smooth. 

Quantities like the Wick square and stress-energy tensor can be defined by normal ordering
relative to some reference state $\omega^R$ using a point-splitting prescription, e.g.,
\[
\langle {:}\Phi^2{:}(x)\rangle_\omega =  (\omega_2-\omega_2^R)(x,x)
\]
One can do without the reference state if, instead of $\omega_2^R$, we subtract a local Hadamard parametrix, i.e., an expression of the form of the RHS of \eqref{eq:Hadamard}, but with $W$ determined
by local geometry rather than a reference state. Actually,
there are remaining freedoms in $W$ that give finite
renormalisation freedoms; we suppose that some choice
has been made and  denote the resulting object by $\Phi^2_\text{ren}$.
The procedure is described in \cite{Wald_qft} with particular reference to the stress-energy tensor;\footnote{As a sketch:  Let ${\mathcal T}$
be a differential operator that maps smooth
functions on $\Mb\times\Mb$ (or local subset thereof) to smooth bi-covector fields,
with the property that
${\mathcal T}(\phi\otimes\phi)(x,x)$ is the classical stress-energy
tensor of any Klein--Gordon solution $\phi$. Applying ${\mathcal T}$ to
$\omega_2-H$, where $H$ is a local Hadamard parametrix, and bringing the points together, we obtain a 
rank-$2$ covariant tensor field $x\mapsto ({\mathcal T}(\omega_2-H))(x,x)$. 
It turns out that although this tensor field is not necessarily conserved, the
problem can be fixed 
by subtracting a local geometrical term of the form $Qg_{ab}$, and can
be avoided altogether by a clever choice of ${\mathcal
T}$~\cite{Moretti:2001}.} see~\cite{Ho&Wa01} for a much more far-reaching development. 
Of course
\[
\langle \Phi^2_\text{ren}(x)\rangle_\omega- \langle \Phi^2_\text{ren}(x)\rangle_{\omega^R} = 
\langle {:}\Phi^2{:}(x)\rangle_\omega 
\]
with normal ordering relative to $\omega^R$.

\subsection{The QEI derivation}\label{sect:QEI_deriv}

Let $\gamma:\RR\to\Mb$ be a smooth timelike curve, with proper time parameterisation. Let $Q$ be any partial differential operator with smooth
real coefficients. We consider the quantity ${:}(Q\Phi)^2{:}$, with normal ordering
performed relative to a reference Hadamard state $\omega^{R}$,
and seek a lower bound on 
\[
\int d\tau\, |g(\tau)|^2 
\langle{:}(Q\Phi)^2{:}\rangle_\omega(\gamma(\tau)) 
\]
for $g\in\CoinX{\RR}$ and any Hadamard state $\omega$. 

To this end, we 
introduce a point-split quantity
\[
G (\tau,\tau') = \langle Q\Phi(\gamma(\tau))Q\Phi(\gamma(\tau'))\rangle_\omega,
\]
and write $G^R$ for the same quantity evaluated in the reference state. 
Both $G$ and $G^R$ are distributions, but their difference $F=G-G^R$ is a smooth
function, which is symmetric [as both $G$ and $G^R$ have equal antisymmetric parts] and whose diagonal $\tau'=\tau$ gives 
\[
F(\tau,\tau) = \langle{:}(Q\Phi){:}^2(\gamma(\tau))\rangle_\omega .
\] 

Then for any real-valued $g\in\CoinX{\RR}$ we compute 
\begin{align*}
\int d\tau\, |g(\tau)|^2 
\langle{:}(Q\Phi)^2{:}\rangle_\omega(\gamma(\tau)) &= 
\int d\tau\, |g(\tau)|^2 F(\tau,\tau)
\\
 & =
\int_{-\infty}^\infty \frac{d\alpha}{2\pi}\int
d\tau\,d\tau'\,g(\tau)
g(\tau') e^{-i\alpha(\tau-\tau')}
F(\tau,\tau') \\
\intertext{(inserting a $\delta$-function to `unsplit' the points)}
&= \int_{-\infty}^\infty \frac{d\alpha}{2\pi} 
F(g_{-\alpha}\otimes g_{\alpha}) \\
\intertext{(thinking of $F$ as a distribution and writing
$g_{\alpha}(\tau)=g(\tau)e^{i\alpha\tau}$)}
&= \int_{0}^\infty \frac{d\alpha}{\pi} F(g_{-\alpha}\otimes g_{\alpha}) 
\end{align*}
using
the symmetry of $F$, and hence $F (g_{-\alpha}\otimes g_{\alpha})=
F(g_{\alpha}\otimes g_{-\alpha})$ to make
the final step. 
As $g$ is real-valued, we have
$g_{-\alpha}(\tau)= \overline{g_{\alpha}(\tau)}$ and obtain
\begin{align}
\int d\tau\, |g(\tau)|^2 
\langle{:}(Q\Phi)^2{:}\rangle_\omega(\gamma(\tau)) &=
\int_{0}^\infty \frac{d\alpha}{\pi} 
F (\overline{g_{\alpha}}\otimes g_{\alpha}) \nonumber
\\
&\ge  -\int_0^\infty
\frac{d\alpha}{\pi} G^R(\overline{g_{\alpha}}\otimes g_{\alpha}), \label{eq:GQEI}
\end{align}
where we have used $F=G-G^R$ and the positive type property of $G$, which it
inherits from $\omega_2$. The positive type property also tells us that
the integrand in the final expression is pointwise positive in $\alpha$.
The result may be generalised to complex-valued $g\in\CoinX{\RR}$ simply
by applying the above argument to the real and imaginary parts separately. 

This derivation provides a quantum inequality on ${:}(Q\Phi)^2{:}$ and
hence on any other quantity that can be expressed as a finite sum of
such quantities. In particular, it applies to the energy density of the minimally coupled  scalar field. Note that the bound depends only on the reference state $\omega_R$ together with $\gamma$ and $g$. We summarise with a theorem
\begin{theorem} 
Let $\Mb$ be any globally hyperbolic spacetime, 
$Q$ be any partial differential operator with smooth real coefficients, $\gamma$
be any smooth timelike curve in a proper-time parameterization. For normal 
ordering performed relative to any Hadamard reference state $\omega^R$, 
the inequality \eqref{eq:GQEI} holds for all Hadamard states $\omega$ of
the real scalar field and all $g\in\CoinX{\RR}$. 
\end{theorem}

However, there are two important questions that must be resolved to complete the
proof of this result:
\begin{itemize}
\item Is it legitimate to restrict the differentiated two-point function to the
world-line, as we did in defining $G$?
\item Is the final integral in \eqref{eq:GQEI} finite? (If not, then the bound would not be
of much interest.) 
\end{itemize}
The (affirmative) answers to these questions require a more in-depth understanding of Hadamard states than we have previously given: namely, using
some tools of microlocal analysis, which are developed in Section~\ref{sect:ulocal}.
However, the reader who does not wish to delve into the details should at least note that
neither is simply a matter of fine precision because
\begin{itemize}
\item the first question would be answered negatively for a null trajectory
and indeed there is no QEI bound in this case~\cite{Fe&Ro03};
\item one may alter the derivation above slightly to yield a bound in which the final integral is taken over the negative
half-line {\em and diverges}.
\end{itemize}

{\noindent\bf Remarks:}
\begin{enumerate}
\item We could equally take averages of other classically positive
contractions of $T_{ab}$ along the timelike curve, e.g., contracted against a null vector or
possibly differing future-pointing causal vector fields, to obtain
QNEI~\cite{Fe&Ro03}, QDEI etc. 
\item Variants exist for averages over suitable Lorentzian submanifolds, instead of
timelike curves (see, e.g.~\cite{FewSmith07}).
\item One may show that no such bounds exist for smearings over spacelike
surfaces (certainly above $2$-dimensions)~\cite{FordHelferRoman:2002} or, 
as already mentioned, along null curves~\cite{Fe&Ro03}. 
\item The argument above, and the analogous argument for the energy
density, relies on `classical positivity' of the quantity in question.
This permits a number of related bounds to be proven by similar methods,
e.g., see \cite{Few&Pfen03} for spin-$1$ fields.  
Nonetheless, there are also QEIs for the free Dirac
field~\cite{FewsterVerch_Dirac,DawsFews06,Smith07} despite the
fact that the `classical' Dirac energy density is symmetrical about zero
and unbounded from below. It turns out that the analogue of the Hadamard
condition also functions as a local version of the Dirac sea, and
restores positivity [modulo a finite QEI lower bound] as well as
renormalising the energy density.
\item See Sect.~\ref{sect:other_dirns} for discussion of
nonminimally coupled scalar fields and the case of interacting QFT.
\end{enumerate}

\paragraph{Dependence on the reference state}
We can rewrite the inequality \eqref{eq:GQEI} as
\[
\int d\tau\, |g(\tau)|^2 
\langle (Q\Phi)^2_{\text{ren}}\rangle_\omega(\gamma(\tau)) 
\ge \int d\tau\, |g(\tau)|^2 
\langle (Q\Phi)^2_{\text{ren}} \rangle_{\omega^R}(\gamma(\tau)) 
 -\int_0^\infty
\frac{d\alpha}{\pi} G^R(\overline{g_{\alpha}}\otimes g_{\alpha})
\]
using a renormalized square, rather than Wick ordering. Now, 
slightly heuristically, $\langle (Q\Phi)^2_{\text{ren}}\rangle_{\omega^R}(\gamma(\tau))$ 
is the diagonal of a function $F_\text{ren}(\tau,\tau')= G^R - G_{\text{ren}}$,
where $G_{\text{ren}}$ is formed from the Hadamard parametrix (i.e., local geometry) and the operator $Q$. So the dependence on the reference state actually cancels, and we obtain
\[
\int d\tau\, |g(\tau)|^2 
\langle (Q\Phi)^2_{\text{ren}}\rangle_\omega(\gamma(\tau)) 
\ge 
 -\int_0^\infty
\frac{d\alpha}{\pi} G_{\text{ren}}(\overline{g_{\alpha}}\otimes g_{\alpha}).
\]
Making this precise and quantitative takes a bit of work~\cite{FewSmith07}.

\subsection{Computations in $n=4$ Minkowski space for minimal coupling}\label{sec:computations}

\paragraph{Inertial trajectory}
Take $Q$ to be a partial differential operator
with constant real coefficients, so that $Qe^{ikx} =p(k)e^{ikx}$
for some polynomial $p$ (which necessarily obeys $p(-k) = \overline{p(k)}$) and adopt the Minkowski vacuum state as the
reference, with 
\[
\omega_2(x,x') = \int \frac{d^3 \kb}{(2\pi)^3} \frac{1}{2\omega}
\text e^{-ik(x-x')} .
\]
With this choice, the normal ordering is precisely the conventional
normal ordering of Minkowski space QFT (and indeed, one would
normally adjust the full renormalized quantity to coincide with this
as well). We take our trajectory to be $\gamma(\tau) = (\tau,\xb)$ in standard inertial coordinates.
Then
\[
G^R(\tau,\tau') = \int \frac{d^3 \kb}{(2\pi)^3} \frac{|p(k)|^2}{2\omega}
\text e^{-i\omega (\tau-\tau')}
\]
and so 
\[
G^R(\overline{g_{\alpha}}\otimes g_{\alpha}) = \int \frac{d^3 \kb}{(2\pi)^3} \frac{|p(k)|^2}{2\omega} \widehat{\overline{g}}(-\omega-\alpha)\widehat{g}(\omega+\alpha)
= \int \frac{d^3 \kb}{(2\pi)^3} \frac{|p(k)|^2}{2\omega} |\widehat{g}(\omega+\alpha)|^2 .
\]

In the specific case of the energy density (of the minimally coupled field) measured along the inertial curve,
the classical expression
\[
T_{00} = \frac{1}{2}\sum_{\mu=0}^3 (\partial_\mu\phi)^2 +\frac{1}{2}m^2
\phi^2
\]
leads us to consider the operators
$Q=2^{-1/2}\partial_\mu$ for $\mu=0,...,3$ and $Q=2^{-1/2}m$. Summing,
\begin{align*}
\int d\tau\, |g(\tau)|^2 
\langle{:}T_{00}{:}\rangle_\omega(\tau,\xb) 
&\ge  -\int_0^\infty
\frac{d\alpha}{\pi}  \int \frac{d^3 \kb}{(2\pi)^3} \frac{\omega^2+\|\kb\|^2+m^2}{4\omega} |\widehat{g}(\omega+\alpha)|^2 \\
&\qquad= -\int_0^\infty
\frac{d\alpha}{\pi}  \int_0^\infty \frac{dk}{4\pi^2} k^2 \omega |\widehat{g}(\omega+\alpha)|^2 \\
\intertext{so changing variables from $k$ to $\omega=\sqrt{k^2-m^2}$,
and then from $(\alpha,\omega)$ to $(\alpha+\omega,\omega)$:}
&\qquad= -\frac{1}{4\pi^3} \int_m^\infty du |\widehat{g}(u)|^2 \int_m^u 
d\omega\, \omega^2 (\omega^2-m^2)^{1/2} \\
&\qquad= -\frac{1}{16\pi^3}\int_m^\infty du |\widehat{g}(u)|^2 u^4 Q_3(u). 
\end{align*}
where $Q_3$ was defined in \eqref{eq:Q3}. This is the bound stated
as \eqref{eq:scalarQI}.

\paragraph{Uniformly accelerated trajectory}
Again we use the vacuum state as the reference, restricting
to massless fields for simplicity. Here it is more convenient to work with the vacuum two-point function in the form
\[
\omega_2(x,x') = \frac{1}{4\pi^2\sigma_+(x,x')}
\]
rather than to use a Fourier representation. We consider
the trajectory 
\[
\gamma(\tau) =(\xi_0\sinh(\tau/\xi_0),\xi_0\cosh(\tau/\xi_0),0,0)
\]
in inertial coordinates,
where $\xi_0>0$ is constant. This is easily seen to be a proper-time
parameterisation of a trajectory with uniform proper acceleration 
$\xi_0^{-1}$.

If we introduce coordinates $x(\tau,\xi,y,z)= (\xi\sinh(\tau/\xi_0),\xi\cosh(\tau/\xi_0),y,z)$
then our trajectory is $\xi=\xi_0$, $y=z=0$. Moreover, on this curve, the vectors
$\partial/\partial\tau$, $\partial/\partial\xi$, $\partial/\partial y$ and $\partial/\partial z$ form
an orthonormal basis and the point-split energy density may be written as 
\[
\frac{1}{2}\left(\frac{\partial^2}{\partial\tau\partial\tau'}+\frac{\partial^2}{\partial\xi\partial\xi'}
+\frac{\partial^2}{\partial y\partial y'}+\frac{\partial^2}{\partial z\partial z'}\right){:}\omega_2{:}(x,x').
\]
In terms of the above coordinates, we have\footnote{A little
justification is needed here, because 
our standard $i\epsilon$ regularisation gives 
\[
\sigma_\epsilon = (\xi-\xi')^2 -4\xi\xi' \sinh^2\left(\frac{\tau-\tau'}{2\xi_0}\right) 
+2i\epsilon \left(\xi\sinh\frac{\tau}{\xi_0} - \xi'\sinh\frac{\tau'}{\xi_0}\right) + \epsilon^2
+(y-y')^2 + (z-z')^2
\]
in these coordinates. The important point is that $\sigma_\epsilon$
has positive imaginary part when $x$ and $x'$ are null-separated
with $x$ to the future of $x'$ (which implies $\tau>\tau'$). This is also true of the corresponding term in our expression for $\omega_2$, 
and so the alternative $i\epsilon$ prescription is valid.}
\[
\omega_2(x,x') = \lim_{\epsilon\to 0+} \frac{1}{4\pi^2}
\left[(\xi-\xi')^2 -4\xi\xi' \sinh^2\left(\frac{\tau-\tau'-2i\epsilon}{2\xi_0}\right) 
+(y-y')^2 + (z-z')^2
\right]^{-1}
\]
and after some calculation one finds that the point-split energy density of the reference state 
restricts to the trajectory as a boundary value distribution
\[
G^R(\tau,\tau')  = T(\tau-\tau'),\quad\text{where}\quad
T(\sigma) = \lim_{\epsilon\to 0+}\frac{3}{32\pi^2\xi_0^4}\cosech^4\left(\frac{\sigma-2i\epsilon}{2\xi_0} \right).
\]
The Fourier transform may be shown to be~\cite{Few&Pfen06}
\[
\widehat{T}(u) = \frac{1}{2\pi}\left(\frac{ u^3 +  u\xi_0^{-2}}{1-e^{-2\pi\xi_0 u}}\right)
\]
and similar calculations to those above give
\begin{equation}
\int d\tau\, |g(\tau)|^2 
\dot{\gamma}^a \dot{\gamma}^b \langle{:}T_{ab}{:}\rangle_\omega(\gamma(\tau)) 
\ge
-\frac{1}{16\pi^3} \int_{-\infty}^\infty du\,|\widehat{g}(u)|^2  \Upsilon(\xi_0,u) 
\label{eq:accQEIa}
\end{equation}
for any Hadamard state $\omega$,
where 
\[
\Upsilon(\xi_0,u)  = 4\int_{-\infty}^u dv\, \frac{ v^3 +  v\xi_0^{-2}}{1-e^{-2\pi\xi_0 v}}.
\]
As $|\widehat{g}(u)|^2$ is even, we may replace $\Upsilon(\xi_0,u)$ by 
$\frac{1}{2}\left(\Upsilon(\xi_0,u)+\Upsilon(\xi_0,-u) \right)$ in \eqref{eq:accQEIa}. 
But one easily sees that
\begin{align*}
\Upsilon(\xi_0,u)+\Upsilon(\xi_0,-u) &= 2\Upsilon(\xi_0,0) + 
4\int_0^u dv\,\frac{ v^3 +  v\xi_0^{-2}}{1-e^{-2\pi\xi_0 v}} 
- 4\int_{-u}^0 dv\,\frac{ v^3 +  v\xi_0^{-2}}{1-e^{-2\pi\xi_0 v}} \\
&= 2\Upsilon(\xi_0,0)  + 
4\int_0^u dv\, ( v^3 +  v\xi_0^{-2})
\underbrace{\left(\frac{1}{1-e^{-2\pi\xi_0 v}} + \frac{1}{1-e^{2\pi\xi_0 v}} \right)}_{=1} \\
&=  u^4 + \frac{u^2}{\xi_0^2} + \frac{11}{30\xi_0^4} ,
\end{align*}
where we have combined the integrals in the second step, changing variables from $u$ to $-u$ in one of them and inserted a closed-form
expression for $\Upsilon(\xi_0,u)$.
Accordingly, the following QEI holds for all Hadamard states $\omega$
and all real-valued $g\in\CoinX{\RR}$:
\begin{align}
\int d\tau\, |g(\tau)|^2 
\dot{\gamma}^a \dot{\gamma}^b \langle{:}T_{ab}{:}\rangle_\omega(\gamma(\tau)) 
&\ge
-\frac{1}{32\pi^3} \int_{-\infty}^\infty du\,|\widehat{g}(u)|^2  \left(u^4 +
 \frac{u^2}{\xi_0^2} + \frac{11}{30\xi_0^4} \right) \notag \\
&\qquad = -\frac{1}{16\pi^2} \int_{-\infty}^\infty d\tau\,\left( |g''(\tau)|^2 + \alpha_0^2 |g'(\tau)|^2 + 
\frac{11\alpha_0^4 }{30}|g(\tau)|^2 \right),
\label{eq:accn}
\end{align}
where $\alpha_0=1/\xi_0$ is the proper acceleration of the trajectory.
Comparing with \eqref{eq:massless4dQEI}, we see that the acceleration 
leads to modifications to the QEI bound that are lower order in the number of derivatives
applied to $g$.

In particular, the scaling behaviour discussed in Section~\ref{sect:exQEI} is modified;
we have
\[
\frac{1}{\tau}\int 
\dot{\gamma}^a \dot{\gamma}^b \langle T_{ab} \rangle_{\omega} (\gamma(t))  |g(t/\tau)|^{2} dt
\ge - \frac{\|g''\|^2}{16{\pi}^{2}\tau^4}- \frac{\|g'\|^2\alpha_0^2}{16{\pi}^{2}\tau^2}- \frac{11\|g\|^2\alpha_0^4}{480{\pi}^{2}}  ,
\]
where the norm is that of $L^2(\RR)$. For $\alpha_0\tau\ll 1$ the previous result is recovered to good approximation;
however, for $\alpha_0\tau\gg 1$ it is the last term that dominates and, indeed, the AWEC fails -- multiplied by $\tau$,
the right-hand side diverges to $-\infty$ as $\tau\to\infty$. By subtracting this troublesome term we can deduce that
\[
\liminf_{\tau\to\infty} \int 
 \left(
\dot{\gamma}^a \dot{\gamma}^b\langle T_{ab} \rangle_{\omega} (\gamma(t))  
+   \frac{11\alpha_0^4}{480{\pi}^{2}}\right) |g(t/\tau)|^{2}\,dt \ge 0
\]
holds for any Hadamard state $\omega$. It is a remarkable fact that the constant negative
contribution is precisely equal to the energy density of the Rindler vacuum state 
for the right-wedge $\{(t,x,y,z): x>|t|\}$ of Minkowski space. (Even though
the Rindler vacuum does not extend to a Hadamard state on the whole of Minkowski
space, it is Hadamard on the interior of the wedge, which completely contains the
accelerated trajectory. Arguments based on local covariance~\cite{BrFrVe03}
show that the Minkowski QEI along that trajectory must be respected by the energy density of
the Rindler vacuum -- see~\cite{Few&Pfen06} for discussion and other examples --
but it is nonetheless surprising that the Rindler vacuum saturates the QEI in this way.)

This example might suggest that a good way of `mining' negative
energy density is simply to follow a uniformly accelerated trajectory,
when the field is prepared in (an approximation to) the Rindler vacuum.
It is worth noting that the work required to maintain this motion grows
exponentially with the proper time, and therefore the `cost' in
work done is growing much more rapidly than the `benefit' of
negative energy `seen'. This seems to fit a broader pattern 
of adverse cost-benefit analyses in other situations where sustained negative energy densities
may be created.

\section{Microlocal analysis and Hadamard states}

\subsection{The wavefront set} \label{sect:ulocal}

Fourier analysis provides a fundamental duality between smoothness and
decay: smooth functions have rapidly decaying Fourier transforms, and
vice versa. The fundamental idea underlying microlocal analysis is that decay
properties of the Fourier transform of a distribution can be used to obtain detailed
information about its singular structure. A general reference for this
section is~\cite{Hormander1}, particularly chapter 8.

\paragraph{The wave-front set} Recall that our convention for the
Fourier transform of functions is  
$$
\widehat{f}(k) = \int d^nx\, e^{ik\cdot x} f(x).
$$
The Fourier transform of a compactly supported distribution $T$ is, 
similarly, $\widehat{T}(k) = T(e_k)$, where $e_k(x) = e^{ik\cdot x}$. 
The Fourier transform of Schwartz distributions can be defined
using duality, because the Fourier transform is an isomorphism of the 
Schwartz space to itself, and hence dually of the Schwartz distributions; 
general distributions in $\DD'$ do not have Fourier transforms.

The duality between smoothness and decay mentioned above
is illustrated by the following examples.
\begin{itemize}
	\item[a)] If $f\in C_0^\infty (\mathbbm R^n)$ then
	$$ 
	\left(1+|k|^{2m} \right) \left|\widehat{f}(k) \right|
	=\left|(1+(-\triangle)^m f)^\wedge (k) \right| \le 
	\int d^n x \left| (1+(-\triangle)^m f)(x)\right| < \infty.
	$$
	So for each $N$, there exists a constant $C_N$ such that
	$$ 
	\left| \widehat{f}(k) \right| \le \frac{C_N}{1+|k|^N} 
	\quad \text { as } k \to \infty \quad 
	$$ 
	(this is what we mean by `rapid decay'.) 
	\item[b)] The $\delta$-distribution at the origin has Fourier
	transform $\widehat{\delta}(k)=1$, which exhibits no decay at $\infty$.
	\item[c)] The distribution $T\in\DD'(\RR)$ defined by 
	$$
	T(f)=\lim_{\varepsilon \to 0^+} \int \frac{f(s)}{s-i\varepsilon} ds
	$$ 
	has Fourier transform
	$$
	\widehat{T}(k)=\lim_{\varepsilon \to 0^+} \int \frac{\text
	e^{iks}}{s-i\varepsilon} ds =2\pi i \Theta(k),
	$$ 
	which decays as $k \to -\infty$ but not as $k \to +\infty$.
\end{itemize}
The wavefront set localises information of this type 
both in $x$-space and on the ``sphere at $\infty$'' in $k$-space.
\begin{definition} (A) 
If $u \in \DD'(\RR^n)$, a pair $(x,k)\in \RR^n \times ((\RR^n)^* \backslash \{0\})$ is a
{\em regular direction} for $u$ if there exist 
\begin{tightenumerate}
	\item[i)] $\phi \in C_0^\infty(\RR^n)$ with $\phi(x)\neq 0$
	\item[ii)] a conic neighbourhood $V$ of $k$ in $(\RR^n)^*$
	\item[iii)] constants $C_N$, $N \in \NN$
\end{tightenumerate}
so that $$\left| \widehat{\phi u}(k) \right|< \frac{C_N}{1+|k|^N} \quad \forall k\in V, N \in \NN$$
i.e., $\widehat{\phi u}$ decays rapidly as $k \to \infty$ in $V$. \\
(B) The {\em wavefront set} of $u$ is defined to be
$$
\WF(u)=
\{(x,k)\in \RR^n\times ((\RR^n)^* \backslash \{0\}): (x,k) \text{ is \emph{not} a regular direction for }u\}.
$$
\end{definition}

%\newpage
{\noindent\em Examples}
\begin{itemize}
	\item[a)] If $f \in C^\infty (\RR^n)$, then $\WF(f)=\emptyset$.
	\item[b)] $\WF(\delta)=\lbrace(0,k) \in \RR^2: k \neq 0\rbrace$.
	(Note that $\widehat{\phi \delta}(k)=\phi(0)$, so $(x,k)$ is a regular
	direction for $x \neq 0$ as we may then choose $\phi$ with $\phi(x) \neq 0$, $\phi(0)=0$).
	\item[c)] $\WF(T)=\lbrace (0,k) \in \RR^2: k > 0\rbrace$ (exercise!).
\end{itemize}

The wavefront set has many natural and useful properties. For our
purposes, the most important are the following: 

\begin{itemize}
	\item $\text{WF}(u)=\phi \Longleftrightarrow u \in C^\infty (\RR^n)$.
	\item $\text{WF} (\lambda u +\mu v) \subset \text{WF}(u) \cup \text{WF}(v)$
for $u,v \in \DD'(\RR^n),\, \lambda,\mu \in \mathbbm C$.
	\item If $P$ is any partial differential operator with smooth coefficients,
then
	$$
	\text{WF}(Pu) \subset \text{WF}(u) \subset \text{WF}(Pu) \cup
	\Char P,
	$$
	for any $u\in\DD'(\RR^n)$, where $\Char P$ is the {\bf characteristic set} of $P$.
To define the characteristic set, let $m$ be the order of $P$, i.e., the
least $m\in\NN_0$ so that $P$ may be written in the form
$P=\sum_{|\alpha|\leq m} a_\alpha(x) (iD)^\alpha$ where $\alpha$ is a
multi-index. The {\bf principal symbol} of $P$ is the smooth function
on $\RR^n\times(\RR^n)^*$ given by
$$
p_m(x,k)=\sum_{|\alpha|=m} a_\alpha(x)
k^\alpha
$$
and the characteristic set is
$$
\Char P=\lbrace (x,k) \in \RR^n \times (\RR^{n*} \setminus \{0\}):
p_m(x,k)=0\rbrace.
$$
	\item {\bf Propagation of Singularities}: 
	$\text{WF}(u) \setminus \text{WF}(Pu)$ is invariant under the Hamiltonian flow generated by $p_m$.
	\item Under coordinate changes, $\text{WF}$ and Char transform as subsets of the cotangent bundle: given a diffeomorphism $\varphi$, define $\varphi^* u$ by $(\varphi^*u)(f)=u(f\circ \varphi)$. Then
	$$ 
	\text{WF}(u)=\lbrace (x,\xi D\varphi|_x): (\varphi(x),\xi) \in \text{WF}(\varphi^*u)\rbrace;
	$$
	similarly, setting $(P_\varphi f)\circ \varphi=P(f \circ \varphi)$, we have
	$$
	(p_\varphi)_m(\varphi(x),\xi)=p_m(x,\xi D\varphi|_x).
	$$
Here $\xi D\varphi|_x$ is the composition of $\xi$ and $D\varphi|_x$ as
linear maps, i.e., the action of the dual map to $D\varphi|_x$ on $\xi$.)
	In particular, we may extend the wavefront set and characteristic
	set to distributions and partial differential operators defined on manifolds;
	both are subsets of the cotangent bundle.
\end{itemize}

{\noindent\em Examples:} 

\noindent 1. Let $P$ be the Klein--Gordon operator
$P=\square_g+m^2+\xi R$ on a spacetime $\Mb$. The
principal symbol is easily seen to be
$$
p_2(x,\xi)=-g^{ab}(x) \xi_a \xi_b
$$
and so the characteristic set is
$$
\Char P=\mathcal{N}:=\lbrace (x,\xi) \in T^* M: \xi \text{ a non-zero null at p}\rbrace ,
$$ 
the bundle of nonzero null covectors on $\Mb$.
Hence the wavefront set of any (distributional) solution to $Pu=0$ obeys
$$
\text{WF}(u) \subset \mathcal N;
$$
moreover, $\text{WF}(u)$ is invariant under the Hamiltonian evolution
$\lambda\mapsto (x(\lambda),\xi(\lambda))\in T^*M$ given by the
`Hamiltonian' $p_2(x,\xi)=-g^{ab}(x) \xi_a \xi_b$. The solution
curve $\lambda\mapsto (x(\lambda),\xi(\lambda))$ is
such that $x(\lambda)$ is a geodesic [which is easily seen by noting that
the `Lagrangian' underlying $p_2$ is $-\frac{1}{4}g_{ab}\dot{x}^a \dot{x}^b$]  
to which $\xi(\lambda)^\sharp$ is
tangent and along which $\xi(\lambda)$ is parallel-transported. 

Recalling that $\WF(u)\subset\mathcal{N}$, we may deduce that 
if $(x,\xi)\in \WF(u)$, then $\xi$ is null, and further, the wavefront set contains every
point $(x(\lambda),\xi(\lambda))$ for $\lambda\in\RR$, where 
$x(\lambda)$ is the null geodesic through $x$ with tangent
$\xi^\sharp$ and $\xi(\lambda)$ is the parallel transport of $\xi$ along
$x(\lambda)$. 

\noindent 2. Now consider Klein--Gordon bisolutions, i.e., $F\in\DD'(M\times M)$ such
that
$$
(P\otimes 1)F = (1\otimes P)F = 0.
$$
Now the operator $P\otimes 1$ has principal symbol
$$
p(x,\xi;x',\xi') = -g^{ab}(x) \xi_a \xi_b
$$
and characteristic set
$$
\Char (P\otimes 1) = (\mathcal{N}_0\times T^*M)\setminus\Zc
$$
where $\mathcal N_0$ is the bundle of (possibly zero) null covectors on
$M$ (i.e., $\mathcal N$, with the zero covector added at each point) and
$\Zc$ is the zero section of $T^*(M\times M)$.  
Similarly, $1\otimes P$ has principal symbol 
$$
p'(x,\xi;x',\xi') = -g^{ab}(x') \xi'_a \xi'_b
$$
and characteristic set
$$
\Char (1\otimes P) = (T^*M\times \mathcal{N}_0) \setminus\Zc
$$
Any bisolution $F$ therefore has wavefront set with upper bound
\begin{align*}
\WF(F) &\subset \Char(1\otimes P)\cap \Char(P\otimes 1) \\
&\subset 
\left(\left(\mathcal{N}_0\times T^*M\right)\cap
\left(T^*M\times \mathcal{N}_0\right) \right))\setminus\Zc \\
&\subset (\mathcal{N}_0\times\mathcal{N}_0)\setminus \Zc
\end{align*}

\paragraph{Pull-backs} Suppose $X$ and $Y$ are smooth
manifolds and $\varphi:Y\to X$ is smooth. Given $u\in\DD'(X)$, 
Theorem 2.5.11${}'$ in~\cite{Hfio1} constructs the pull-back
$\varphi^*u$ as a distribution on $Y$ provided 
$\WF(u)\cap N_\varphi=\emptyset$, where 
\begin{equation}
N_\varphi = \left\{(\varphi(y),\xi)\in T^*X\mid
\xi D\varphi(y)=0\right\} 
\end{equation}
defines the set of normals of the map $\varphi$. The wave front set of
the pull-back is constrained by
\begin{equation} 
\WF(\varphi^*u) \subset \varphi^* \WF(u)=
\left\{(y, \xi D\varphi(y))\mid (\varphi(y),\xi)\in \WF(u)\right\}\,.
\label{eq:pullWF}
\end{equation}
If $u$ is smooth, the pull-back reduces to ordinary composition
$\varphi^*u(y)=u(\varphi(y))$.

{\noindent\em Example} Let $F\in\DD'(\Mb\times\Mb)$ be 
a Klein--Gordon bisolution, and let $\gamma:\RR\to\Mb$ be a smooth timelike curve.
We wish to consider the pull-back $(\gamma\times\gamma)^*F$. 

To see that this is well-defined, we first compute the set of normals to 
$\gamma\times\gamma:\RR^2\to\Mb\times\Mb$, noting that
\[
\begin{pmatrix} k & k'\end{pmatrix}
D(\gamma\times\gamma)|_{(\tau,\tau')}\begin{pmatrix} t \\ t'\end{pmatrix}
= (\dot{\gamma}(\tau)\cdot k) t + (\dot{\gamma}(\tau')\cdot k') t'
\]
and therefore vanishes for all $t,t'$ iff $\dot{\gamma}(\tau)\cdot k = \dot{\gamma}(\tau')\cdot k' =0$.
Thus
\[
N_{\gamma\times\gamma} = \left\{(\gamma(\tau),k;\gamma(\tau'),k') \in T^*M\times T^*M
\mid \dot{\gamma}(\tau)\cdot k = \dot{\gamma}(\tau')\cdot k' =0\right\} .
\]
Now the covectors arising in $\WF(F)$ are always null and at least one of them must be
nonzero; moreover, no nonzero null covector
can have vanishing contraction with a timelike vector. Thus
\[
\WF(F) \cap N_{\gamma\times\gamma}  =\emptyset
\]
and the pull-back is well-defined, with wave-front set
\begin{align*}
\WF(G) &\subset (\gamma\times\gamma)^* \WF(F) \\ &=
\left\{(\tau,\dot{\gamma}(\tau)\cdot k;\tau',\dot{\gamma}(\tau')\cdot k')\in T^*\RR\times T^*\RR \mid 
(\gamma(\tau),k;\gamma(\tau'),k') \in \WF(F)\right\} .
\end{align*}
The same is true for any distribution $QF$, where $Q$ is a partial differential operator on $\Mb\times\Mb$ with smooth coefficients, because the
wave-front set cannot expand under the action of $Q$.

There are similar wave-front set conditions under which products of distributions can
be defined. 

\subsection{Microlocal formulation of the Hadamard condition}\label{sect:uSC}

Let us compute the wave-front set of the Minkowski vacuum $2$-point function 
\[
\omega_2(x,x') = \int \frac{d^3 \kb}{(2\pi)^3} \frac{1}{2\omega}
\text e^{-ik(x-x')}.
\]
Consider a localising function of the form $\phi(x_1,x_2)=\phi_1(x_1)\phi_2(x_2)$ where $\phi_i \in
C_0^\infty(M)$. Then
$$
\widehat{\phi \omega_2} (\ell,\ell')=
\int \frac{d^3\kb}{(2\pi)^3} \frac{1}{2\omega} \widehat{\phi}_1(\ell-k)
\widehat{\phi}_2(\ell'+k)
$$
with future pointing, on-shell $k$. As the functions $\phi_i$ are
smooth, their Fourier transforms decay rapidly as their arguments become
large. The main contribution to the integral therefore arises from regions of $k$ where $\ell-k$
and $\ell'+k$ are simultaneously small, i.e., $\ell$ must be near to the
future pointing on-shell covector $k$, and $\ell'$ must be near $-k$.
Arguing in this way, it is not hard to see that there are open
conic neighbourhoods of $\RR^4_-\times \RR^4$ and
$\RR^4\times \RR^4_+$ in which the integral will
tend rapidly to zero as $(\ell,\ell')\to\infty$, 
where $\RR^4_\pm$ is the half-space in which
$\pm k_0\ge 0$.

\begin{figure}[th]
\begin{center}
\begin{tikzpicture}%[domain=-1:0.5]
%\draw[dashed] (0,4) -- ++(6,0);
\draw[smooth,domain=-4:4,line width=1.5pt]  plot(\x,{sqrt(abs(\x)^2+4)})
node[above]{$k^2=m^2$};
\draw[dashed] (-4,4) -- (0,0) -- (4,4);
\def\mypath{(-4,1) -- (0,0) -- (4,1) -- (4,-2) -- (-4,-2) -- (-4,1)};
\fill[pattern = north west lines] \mypath;
\draw[very thin, ->] (0,-2) -- (0,5) node[right] {$k_0$};
\draw[very thin, ->] (-4,0) -- (4.2,0) node[right] {$\kb$};
%\node[below left] at (0,0) {$0$};
%\node[below] at (1,0) {$1$};
%\node[left] at (0,4) {$1$};
\end{tikzpicture}
\end{center}
\caption{If $\ell\to\infty$ in the shaded region then $\widehat{\phi \omega_2} (l,l')\to 0$ rapidly, 
regardless of $\ell'$.}
\end{figure}
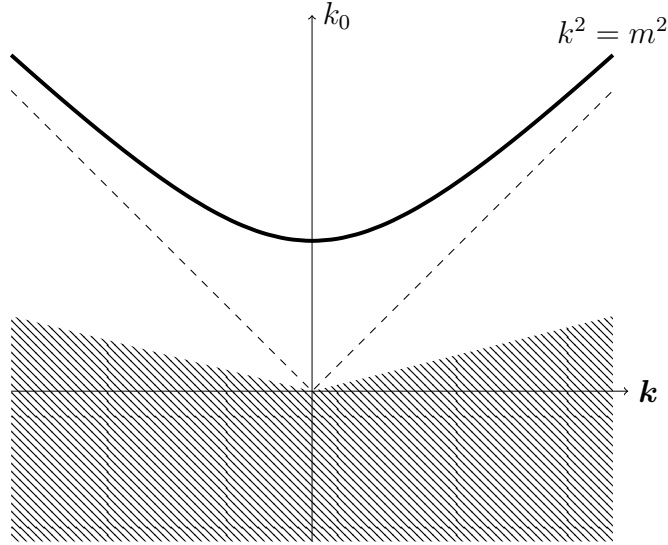

Thus $(x_1,k_1;x_2,k_2)$ is a
regular direction if either (i) $(k_1)_0\le 0$ or (ii) $(k_2)_0\ge 0$.
Putting this together with the upper bound
\[
\WF(\omega_2) \subset \mathcal N_0 \times \mathcal N_0
\] 
we conclude that
\begin{equation}\label{eq:WF_Minkowski_vacuum}
\WF(\omega_2)\subset \mathcal N^+ \times \mathcal N^-,
\end{equation}
where
$$
\mathcal N^\pm=\lbrace (p,\xi) \in \mathcal N: \xi \text{ is future($+$)/past($-$) directed} \rbrace.
$$ 
Now any Hadamard state of the Minkowski theory must have the same $2$-point wave-front set, because $2$-point functions
of Hadamard states differ by smooth functions. 
We now elevate this to a general principle in curved spacetimes. 

\begin{definition} A state $\omega$ obeys the {\em\bf Microlocal Spectrum
Condition} ($\mu$SC)\footnote{The term microlocal spectrum condition was
introduced in~\cite{BFK:1995} with an apparently stronger definition;
see the remarks at the end of this section.} if 
$$
\WF(\omega_2)\subset \mathcal N^+ \times \mathcal N^-.
$$
\end{definition}

In particular, this asserts that the `singular behaviour' of the
two-point function is positive-frequency in the first slot and
negative-frequency in the second. We have already argued that the
Minkowski vacuum obeys the $\mu$SC; it is also true that ground and
thermal states on various classes of stationary spacetime satisfy
the $\mu$SC 
\cite{Junker:1996,SahlmannVerch_passivity:2000,SVW2002}. 
(In relation to thermal states, the key point is that 
negative frequency contributions to the first slot of the
thermal two-point functions are exponentially suppressed,
rather than being absent. But this is enough to get the
necessary decay properties.)

A truly remarkable fact is that the $\mu$SC is enough to completely 
fix the singular structure of $\omega_2$ (and even more: see
remarks at the end of this section). Bear in mind that
even if two distributions have the same wavefront
set, their difference is not necessarily smooth
($\WF(\delta)=\WF(2\delta)$, for instance). The following
result is due to Radzikowski~\cite{Radzikowski_ulocal1996}.
\begin{theorem} \label{thm:Radz}
If $\omega$ and $\omega'$ obey the $\mu$SC then
\[
\omega_2-\omega'_2 \in C^\infty(M\times M)
\]
i.e., the $\mu$SC determines an equivalence of class of states under
equality of two-point functions modulo $C^\infty$. Moreover, 
the $\mu$SC is equivalent to the Hadamard condition. 
\end{theorem}

As mentioned, it is surprising that such a result can be true. 
The key point is that, while the antisymmetric parts of $\omega_2$ and
$\omega_2'$ are both equal to $\frac{1}{2}i E$, $\text{WF}(\omega_2)$
is not the whole of $\text{WF}(E)$,
which intersects both $\mathcal N^+ \times \mathcal N^-$ and $\mathcal N^- \times \mathcal
N^+$. Accordingly, the singularities in the symmetric
part must precisely cancel the unwanted singular directions in $\WF(E)$, which is
how the microlocal spectrum condition does, after all, fix the
singular structure of the two-point function.  

It follows from Theorem~\ref{thm:Radz} that all two-point
functions of states obeying $\mu$SC must have equal wavefront sets. The
universal nature of the antisymmetric part of the two-point function 
also allows us to fix the wavefront set of the two-point
function as follows. 
\begin{lemma} If $\omega$ obeys the $\mu$SC then
$\WF(\omega_2)=\WF(E) \cap (\mathcal N^+ \times \mathcal N^-)$.	
\end{lemma}
{\noindent\em Proof:} Define $\tilde \omega_2(x,x')=\omega_2(x',x)$, so 
$\WF(\tilde \omega_2) \subset \mathcal N^- \times \mathcal N^+$ by
the $\mu$SC. But, using successively that $iE=\omega_2-\tilde \omega_2$ and
$\omega_2=\tilde \omega_2 +iE$, we find
\[
\WF(E) \subset \WF(\omega_2) \cup \WF(\tilde \omega_2) \subset \WF(\tilde \omega_2) \cup \WF(E) 
\]
so, using again the fact that $\WF(\tilde \omega_2) \subset \mathcal N^- \times \mathcal N^+$
$$ 
\WF(E) \subset \text{WF}(\omega_2) \cup (\mathcal N^- \times \mathcal N^+)
\subset (\mathcal N^- \times \mathcal N^+) \cup
\WF(E),
$$
and we take intersections with $\mathcal N^+ \times \mathcal N^-$
to obtain the required result. $\square$

The wavefront set of $E$ is known from work of Duistermaat and
H\"ormander on distinguished parametrices. This permits us to give a
final form of the wavefront set of a Hadamard $2$-point function:
\begin{equation}\label{eq:WFSSC}
\WF(\omega_2) = \{ (p,\xi;p, -\xi')\in T^*(M\times M)\setminus\Zc:
~(p,\xi)\sim(p',\xi')~\text{and}~\xi\in{\mathcal N}^+\},
\end{equation}
where the equivalence relation $\sim$ is defined so that
$(p,\xi)\sim (p',\xi')$ if and only if either
\begin{tightitemize}
\item there is a null geodesic $\gamma$ connecting $p$ and $p'$, so that
$\xi$ is parallel to $\dot{\gamma}^\flat$ at $p$, and $\xi'$ is the
parallel transport of $\xi$ to $p'$ (and necessarily parallel to
$\dot{\gamma}^\flat$ at $p'$); or,
\item $p=p'$ and $\xi=\xi'$.
\end{tightitemize}
Eq.~\eqref{eq:WFSSC} is the form that Radzikowski stated as his
`wavefront set spectral condition'. 

\medskip
{\noindent\bf Remarks:}
\begin{enumerate} 
%\item As mentioned above, the final stress-energy
%tensor exhibits a {\bf trace anomaly}: even if the classical theory is
%conformally invariant (e.g., $m=0$ and $\xi=1/6$ in four dimensions) the
%trace of the renormalised stress-energy tensor is nonvanishing, in
%contrast to the classical situation. 
%\item The length scale $\ell$ provides a residual freedom
%in the definition of $H_\OO$ and hence a one-parameter family of renormalized Wick
%squares and stress-energy tensors. In the case of the stress-energy
%tensor there is a wider four-parameter family of alternatives, differing
%by conserved symmetric local curvature tensors that are at most
%quadratic in the curvature, which obey axioms for
%renormalization due to Wald and arise in other renormalization schemes.
%(See~\cite{Wald_qft}; the original
%references \cite{Wald_axioms:1977,Wald_traceanomaly:1978} include an additional axiom later
%dropped from the list.) For similar issues in the case of higher Wick
%powers see \cite{Ho&Wa01}, and \cite{BuchSchlem07} for a
%proposal to fix these freedoms using thermodynamic data). 
%These freedoms are often called ambiguities, but in my view are better
%regarded as relating to different extensions of the same underlying free
%theory, distinguished by the renormalized products of
%the underlying fields.
\item As the following quotation, taken from the 1978 paper of Fulling,
Sweeny and Wald \cite{FullingSweenyWald:1978},
makes clear, the introduction of the Hadamard condition was a spur to the
development of the algebraic approach to QFT in CST: 
\begin{quote}
All these considerations suggest that the validity of [the Hadamard
condition] be regarded as a basic
criterion for a ``physically reasonable'' state, perhaps even as the definition of that
phrase. This raises the possibility of constructing quantum states from two-point
distribution solutions of the field equation by a procedure of the Wightman or
GNS type... ...bypassing the quantization of normal modes in a Fock space.
\end{quote}
\item We have only discussed regularity of the $2$-point function. In
some references, the term microlocal spectrum condition is defined as a
condition on all $n$-point functions of the form
$$
\WF(\omega_n)\subset \Gamma_n,
$$
where the $\Gamma_n$ are particular subsets of $T^*M^{\times n}$. This condition was
introduced in~\cite{BFK:1995}, where it is also shown to be satisfied by
all quasifree Hadamard states. Very recently, Sanders has
proved that this apparently more general condition is actually
equivalent to the $\mu$SC in the
form we have stated; and, moreover, that all states obeying $\mu$SC
have smooth truncated $n$-point functions for
$n\not=2$~\cite{Sanders_thesis,Sanders:2010}. One may
interpret this as saying that all Hadamard states are `microlocally
quasifree'; it also shows that the class of (not necessarily quasifree)
Hadamard states is precisely
the `state space of perturbative QFT' studied by Hollands \& Ruan in~\cite{HollandsRuan},
and previously identified as a plausible class of interest by Kay~\cite{Kay:1991qx}. 
\item Finally, we mention a variation on the theme. For some purposes it
is sufficient only to require the two-point
functions to agree with a Hadamard parametrix modulo corrections in some Sobolev space, rather than
modulo $C^\infty$. This leads to the microlocal study of {\bf adiabatic states}
\cite{JunkerSchrohe2001}. 
\end{enumerate}

\subsection{Application to QEIs}\label{sect:uloc_QEI_applic}

There were two issues to resolve in completing the proof of the general QEI in
Sect.~\ref{sect:QEI_deriv}. The first was to establish the validity of restricting the differentiated $2$-point function to the worldline. Effectively
we want to define
\[
G(\tau,\tau') = \left((Q\otimes Q)\omega_2)\right)(\gamma(\tau),\gamma(\tau')) ,
\]
i.e., $G =  (\gamma\times\gamma)^* \left((Q\otimes Q)\omega_2)\right)$. 
This is well-defined by the example at the end of Section~\ref{sect:ulocal};
moreover, 
\begin{align*}
\WF(G) &=
\left\{(\tau,\dot{\gamma}(\tau)\cdot k;\tau',\dot{\gamma}(\tau')\cdot k')\in T^*\RR\times T^*\RR \mid 
(\gamma(\tau),k;\gamma(\tau'),k') \in \WF(\omega_2)\right\} \\
&\subset \RR\times\RR^+\times\RR\times\RR^-,
\end{align*}
because the covector in the first slot of $\WF(\omega_2)$ is future-pointing,
as is $\dot{\gamma}(\tau)$, while the covector in the second slot is past-pointing.
Here we have used both the Hadamard condition and the timelike nature of the curve in an
essential way. The same results apply to $G^R$, of course. 

The second question concerned the convergence of
\[
\int_0^\infty
\frac{d\alpha}{\pi} G^R(\overline{g_{\alpha}}\otimes g_{\alpha}).
\]
Now the integrand is
\[
\int d\tau\,d\tau' g(\tau)e^{-i\alpha\tau} g(\tau')e^{i\alpha\tau'} G^R(\tau,\tau') = 
[(g\otimes g)G^R]^\wedge(-\alpha,\alpha)
\]
and this decays rapidly as $\alpha\to+\infty$ by definition of the wave-front set,
and the bound $\WF(G^R) \subset \RR\times\RR^+\times\RR\times\RR^-$. Thus we have convergence of the integral and a finite bound -- and we also see why it would have been a bad idea to arrange the final integral in terms
of an integral over the negative half-line. 

\section{Conformal field theories}

\subsection{Derivation of the QEI}\label{sect:CFT_QEI}

Conformal quantum field theories in two-dimensions provide examples of non-free fields for which
quantum inequality results may be derived. The basic idea was given by Flanagan~\cite{Flanagan:1997} for massless scalar fields. It was generalised to massless Dirac fields
by Vollick~\cite{Vollick:2000} and made into a general and rigorous argument for CFTs by Fewster \& Hollands~\cite{Fe&Ho05}. We will not emphasize analytical details here, although everything can be made precise and rigorous. Throughout, we work in two-dimensional Minkowski space; a general
reference is~\cite{FurlanSotkovTodorov:1989}.

Recall that the stress tensor $T^{ab}$ in CFT is traceless and splits into chiral components
\begin{eqnarray*}
T^{00}(t,x) &=& T_R(t-x) + T_L(t+x)\\
T^{01}(t,x) &=& T_R(t-x) - T_L(t+x)
\end{eqnarray*}
and that the left- and right-moving chiral components 
$T_L$ and $T_R$ commute and obey the spectrum condition
\[
P_{L,R}=\int T_{L,R}(v)\,dv \ge 0 .
\]

The important feature of CFTs we will use is that
reparameterisations of null coordinates $v=t-x$, $u=t+x$ are unitarily
implemented, in the following sense. Under the correspondence
$v\mapsto z(v) := (i-v)/(i+v)$, the real-line is mapped to
$\mathbb{T}\setminus\{-1\}$, where $\mathbb{T}\cong S^1$
is the unit circle in $\CC$. If a reparameterisation $v\mapsto V(v)$ lifts to an orientation preserving diffeomorphism of $S^1$, then there is a unitary
$U_R(V)$ s.t.
\[
U_R(V)T_R(v)U_R(V)^{-1} = V'(v)^2 T_R(V(v)) -
\frac{c_R}{24\pi}\{V,v\}{\boldsymbol{1}},
\]
where $ c_R$ is the central charge (for right-movers) and
\[
\{V,v\} = -2\sqrt{V'(v)}\frac{d^2}{dv^2}\frac{1}{\sqrt{V'(v)}}
\]
is the Schwarzian derivative. The same is true for $T_L$ and the unitaries $U_L$, $U_R$ commute.\footnote{More generally, the theory contains commuting `left' and `right' unitary multiplier representations of the universal covering group of the orientation-preserving diffeomorphisms of
$S^1$, obeying
\[
U(\rho)U(\rho') = e^{icB(\rho,\rho')}U(\rho\rho')
\]
where $B$ is the Bott cocycle and $c$ is the central charge.}

Consider one of the stress-tensor components, say, $T(v)$ and let $f$ be a smooth
compactly supported positive real-valued function. We define
\[
T(f) = \int T(v)f(v)\,dv
\]
and aim to show that there is a lower bound on the expectation values $\langle T(f)\rangle_\psi$. 

The idea is to define $v\mapsto V(v)$ by $V'(v)=1/f(v)$ and set $\psi'=U(V)\psi$. Then
\begin{align*}
f(v)\langle T(v)\rangle_\psi & = f(v) \langle U(V)T(v)U(V)^{-1}\rangle_{\psi'} \\
&=V'(v)\langle T(V(v))\rangle_{\psi'}
-\frac{c}{24\pi}\{V,v\}f(v) 
\end{align*}
so 
\begin{eqnarray*}
\int \langle T(v)\rangle_\psi f(v)\,dv &=& 
\int \langle
T(V)\rangle_{\psi'}\,dV- 
\frac{c}{24\pi}\int \{V,v\} f(v)\,dv \\
&=& \langle P\rangle_{\psi'} - \frac{c}{24\pi}\int \{V,v\} f(v)\,dv.
\end{eqnarray*}
Using the spectrum condition, and rearranging, we find
\[
\int \langle T(v)\rangle_\psi f(v)\,dv \ge  - \frac{c}{12\pi}\int \left(\frac{d}{dv}\sqrt{f(v)}\right)^2\,dv
\]
for all `reasonable' $\psi$. 

The only problem is that the map $v\mapsto V(v)$ does not lift to a diffeomorphism  of $S^1$. The  resolution is to replace $f$ by 
\[
f_{n\epsilon}(v)=f(v)+\epsilon -
\zeta_{n\epsilon}(v)\]
with $\zeta_{n\epsilon}$ chosen so that
\begin{tightitemize}
\item $V'(v)=1/f_{n\epsilon}(v)$ does lift to a diffeomorphism of $S^1$
for all $n\in\NN$, $\epsilon>0$;
\item $\displaystyle\int \langle T(v)\rangle_\psi
\zeta_{n\epsilon}(v)\,dv\longrightarrow 0$ as $n\to\infty$ for each
fixed $\psi,\epsilon$;
\item 
$\displaystyle\lim_{\epsilon\to \smash{0^+}}\lim_{n\to\infty}
\int\left(\frac{d}{dv}\sqrt{f_{n\epsilon}(v)}\right)^2\,dv = \int  
\left(\frac{d}{dv}\sqrt{f(v)}\right)^2\,dv$.
\end{tightitemize}
Then, for each fixed $\psi$
\[
\int \langle T(v)\rangle_\psi f_{n\epsilon}(v)\,dv \ge  - \frac{c}{12\pi}\int
\left(\frac{d}{dv}\sqrt{f_{n\epsilon}(v)}\right)^2\,dv
\]
so taking $n\to\infty$...
\[
\int \langle T(v)\rangle_\psi f(v)\,dv +\epsilon\langle P\rangle_\psi \ge  - \frac{c}{12\pi}\lim_{n\to\infty}\int
\left(\frac{d}{dv}\sqrt{f_{n\epsilon}(v)}\right)^2\,dv
\]
...and $\epsilon\to 0^+$, we obtain the desired bound
\[
\int \langle T(v)\rangle_\psi f(v)\,dv \ge  - \frac{c}{12\pi}\int \left(\frac{d}{dv}\sqrt{f(v)}\right)^2\,dv\,.
\]

The fully rigorous argument for this is given in \cite{Fe&Ho05}, where 
an axiomatic setting is adopted in which all the above manipulations are
justified and the class of `reasonable' $\psi$ is specified. The axioms are shown to
hold for CFTs constructed from unitary, positive
energy Virasoro representations. We also proved that the bound is sharp if the theory has a conformally invariant vacuum. Any nonnegative $f\in\mathscr{S}(\RR)$ can be used for smearing.

This argument is notable, partly as the first examples of QEIs for non-free fields,
but also because it does not depend on a `sum of squares' form of the energy
density. It is also model-independent, applying to {\em all} unitary positive energy CFTs in one go.

\subsection{Probability distributions}

Everything said so far concerns the expectation value of the smeared stress-energy tensor
or other similar quantities. Here, we discuss what information can be gleaned concerning
the underlying probability distribution of individual measurements of such quantities. Again, CFTs provide a framework in which this
can be studied for a whole class of models. The argument given here is taken from~\cite{FewsterFordRoman:2010} and approaches  the probability distribution through its moment generating function
\[
M[\mu f] = \sum_{n=0}^\infty \frac{\mu^n\ip{\Omega}{T(f)^n\Omega}}{n!}.
\]

Our notation is
\begin{align*}
G_n(u_n,\ldots,u_1) &= \ip{\Omega}{ T(u_n)\cdots T(u_1)\Omega} \qquad\qquad (G_0=1)\\
\GG_n[f] &=
G_n(f,\ldots,f) = \ip{\Omega}{T(f)^n\Omega}
\end{align*}
and we assume that test functions are real-valued and rapidly decaying at infinity.
The main tool used in the argument is the CFT Ward identity~\cite[p.~28]{FurlanSotkovTodorov:1989}\footnote{Beware, however, a misprint in Eq.~(3.12a) of \cite{FurlanSotkovTodorov:1989} [$\Theta^{(-)}$ should be $\Theta^{(+)}$]. Fortunately the result given before (3.15) of \cite{FurlanSotkovTodorov:1989} is correct.} 
\begin{align*}
G_n(u_n,\ldots,u_1) &= \sum_{j=1}^{n-1}\left[
\frac{c}{8\pi^2}\frac{G_{n-2}(u_{n-1},\ldots,
\hat{u}_j,\ldots,u_1)}{(u_n-u_j-i0)^4}\right. \\
 &\qquad-\frac{\partial_j G_{n-1}(u_{n-1},\ldots,
u_1)}{2\pi(u_n-u_j-i0)} \left.
-\frac{G_{n-1}(u_{n-1},\ldots,u_1)}{\pi(u_n-u_j-i0)^2}\right],
\end{align*}
where the hat denotes an omitted variable. 

Since $G_0=1$ and
$G_1(u_1)=\langle T(u_1)\rangle \equiv 0$, it follows immediately that
\begin{equation}
G_2(u_2,u_1) = \frac{c}{8\pi^2(u_2-u_1-i0)^4}
\end{equation}
and if we smear the Ward identity against $n$ copies of $f$, we find
\[
\GG_n[f] = (n-1) \GG_2[f]\GG_{n-2}[f] + \sum_{j=1}^{n-1} I_{n,j} \,,
\]
where 
\begin{align*}
I_{n,j} &=-\frac{1}{2\pi} \int du_n\,du_j\, f(u_n)f(u_j) \left[
\frac{1}{u_n-u_j-i0}\partial_j G_{n-1}(f,\ldots,u_j,\ldots,f)\right.\\
&\qquad\left.
+\frac{2}{(u_n-u_j-i0)^2}G_{n-1}(f,\ldots,u_j,\ldots,f)\right] 
\\&= G_{n-1}(f,\ldots,\underbrace{f\star f}_j,\ldots,f)
\end{align*}
after integration by parts. Here, $f\star f$ is 
\begin{eqnarray}
f\star f(u_j) &=& \int du_n\,
f(u_n)\left[\partial_{j}\left(\frac{f(u_j)}{2\pi(u_n-u_j-i0)}\right)-
\frac{f(u_j)}{\pi(u_n-u_j-i0)^2}\right]\nonumber\\
&=&\int du_n\, \left( \frac{f(u_n)f'(u_j)}{2\pi(u_n-u_j-i0)}-
\frac{f(u_n)f(u_j)}{2\pi(u_n-u_j-i0)^2}\right)\nonumber\\
&=&\int du_n\,\left( \frac{f(u_n)f'(u_j)}{2\pi(u_n-u_j-i0)} +
f(u_n)f(u_j)\partial_{n}\frac{1}{2\pi(u_n-u_j-i0)}\right)\nonumber\\
&=& \int du_n\, \frac{f(u_n)f'(u_j)-f'(u_n)f(u_j)}{2\pi(u_n-u_j-i0)} \\
&=& \int du_n\, \frac{f(u_n)f'(u_j)-f'(u_n)f(u_j)}{2\pi(u_n-u_j)}
\end{eqnarray}
after using the Leibniz rule, a further integration by parts in one
term and observing that the the numerator in the penultimate integrand
vanishes as $O(u_n-u_j)$ as $u_n\to u_j$.
Note that no boundary terms arise when integrating by parts provided $f$ is compactly supported,
for instance.

To solve the recurrence relation, 
consider a $1$-parameter family of test functions $(f_\lambda(u))_{\lambda\in\RR}$
solving 
\[
\frac{df_\lambda}{d\lambda} = f_\lambda\star f_\lambda \qquad f_0=f.
\]
Then the recurrence relation becomes
\[
\GG_n[f_\lambda] = (n-1)\GG_2[f_\lambda]\GG_{n-2}[f_\lambda]+
\frac{d}{d\lambda}\GG_{n-1}[f_\lambda]
\]
and gives a p.d.e.
\[
\frac{\partial}{\partial \mu} M[\mu f_\lambda]
= \mu\GG_2[f_\lambda] M[\mu f_\lambda] + 
\frac{\partial}{\partial\lambda} M[\mu f_\lambda] 
\]
using $\GG_0[f_\lambda]=1$, $\GG_1[f_\lambda]=0$. 
Solving, using the fact that $M[\mu f_\lambda]|_{\mu=0} = 1$ and setting $\lambda=0$, 
\[
M[\mu f] = \exp\left(\int_0^\mu d\lambda\, (\mu-\lambda)\GG_{2}[f_{\lambda}]\right),
\]
a result first obtained by Haba~\cite{Haba:1990}. 

In general, it is not easy to take this further. However, if $f$ is Gaussian, we may proceed to a 
closed form result. Let $f(u) = e^{-u^2/\tau^2}/(
\tau\sqrt{\pi})$ and make an {\em ansatz} 
$f_\lambda(u) = A(\lambda)f(u)$. Then
\[
(f_\lambda\star f_\lambda)(u) =
\frac{A(\lambda)^2}{\tau^3\pi^{3/2}}e^{-u^2/\tau^2} =
\frac{A(\lambda)}{\tau^2\pi} f_\lambda(u),
\]
so the flow equation for $f_\lambda$ reduces to
\[
A'(\lambda) =  \frac{A(\lambda)^2}{\tau^2\pi}.
\]
The unique solution with $A(0)=1$ is
\[
A(\lambda) = \frac{\pi\tau^2}{\pi\tau^2-\lambda}.
\]
Thus $\GG_2[f_\lambda]=A(\lambda)^2\GG_2[f]$ and we calculate
\[
M[\mu f] =
\left[\frac{e^{-\mu/(\pi\tau^2)}}
{1-\mu/(\pi\tau^2)}\right]^{c/24}.
\]

The probability distribution itself is then obtained essentially by inverse Laplace
transformation: we seek $P(\omega)$ such that
\[
M[\mu f] = \int_{-\infty}^\infty d\omega\, P(\omega) e^{\mu\omega}.
\]
and the solution is
\[
P(\omega) =
\vartheta(\omega+\omega_0) 
\frac{\beta^{\alpha}(\omega+\omega_0)^{\alpha-1}}{\Gamma(\alpha)} 
\exp(-\beta(\omega+\omega_0)),
\]
(a shifted Gamma distribution) with parameters 
\[
\omega_0 = \frac{c}{24\pi\tau^2},\qquad
\alpha = \frac{c}{24}, \qquad
\beta = \pi\tau^2,
\]
which has an integrable singularity at lower limit for $c<24$. 

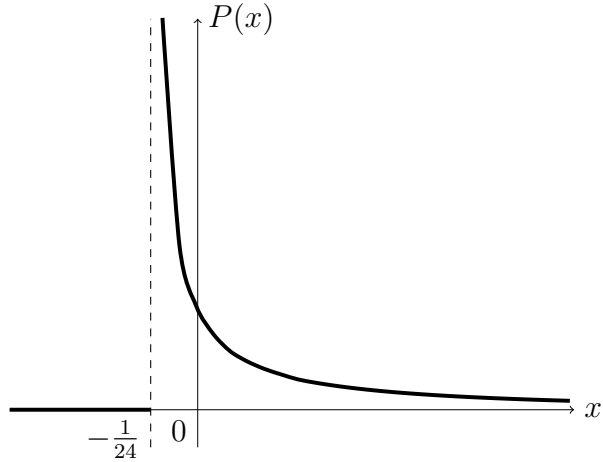
\begin{figure}[th]
\begin{center}
\begin{tikzpicture}%[domain=-1:0.5]
\draw[very thin, ->] (-2.5,0) -- (5,0) node[right] {$x$};
\draw[very thin, ->] (0,-0.5) -- (0,5.2) node[right] {$P(x)$} ;
\draw[dashed] (-15/24,-0.5) -- ++(0,5.7);
%  plot coordinates{(1,1) (2,2) (3,3)};
\draw[smooth,domain=-0.75/24:0.33,line width=1.5pt] plot (15*\x,{(\x+1/24)^(-23/24)*exp(-\x-1/24)/15});
\draw[line width =1.5pt] (-15/24,0) -- (-2.5,0);
\node[below left] at (0,0) {$0$};
%\node[above left] at (0,2) {$1$};
\node[below left] at (-15/24,0) {$-\frac{1}{24}$};
\end{tikzpicture}
\end{center}
\caption{The probability density $P(x)$ plotted for $c=1$, where $x=\pi\tau^2\omega$}
\end{figure}

One should mention that the Hamburger moment theorem guarantees that this
is the only possible solution: noting that it is {\em a} solution, we may use it
to read off the moments $a_n$ of $P(\omega)$ and note that they obey a bound
$|\GG_n[f]|\le BC^n n!$ for some constants $B$, $C$, thereby satisfying the hypotheses of the Hamburger uniqueness theorem~\cite{Simon:1998}.

The probability distribution is clearly highly skewed. We see that the lower bound
of the support coincides precisely with the sharp lower bound on the expectation values of $T(f)$ for the Gaussian $f$, i.e., $-c/(24\pi\tau^2)$ -- as it should for general reasons~\cite{FewsterFordRoman:2010}. Thus the QEI bound,
which was originally derived as a constraint on the {\em expectation value} of 
the smeared energy density, also turns out to be a constraint on the minimum
value that can be achieved in an {\em individual measurement} of this quantity.

The probability of obtaining a negative value is given in terms of incomplete $\Gamma$-functions:
\[
\text{Prob}(\omega<0) = 1- \frac{\Gamma(c/24,c/24)}{\Gamma(c/24)}.
\]
For $c=1$, this results in a value $0.89$ -- an overwhelming likelihood of
obtaining a {\em negative} value from a measurement in the vacuum state.
In the limit $c\to\infty$ the probability tends to $1/2$, to be expected from
the central limit theorem. (Note: the above computation
refers to just one of the chiral components of $T_{ab}$, and
for averaging along the corresponding light-ray. For Gaussian averages
of the energy density along an inertial curve and $c=1$,
the probability of obtaining a negative value is $0.84$.)

It is somewhat ironic that negative energy densities, which are suppressed (in
all physically reasonable states) by the uncertainty principle expressed in the QEIs, turn out to occur with such high probability, in individual measurements made in the {\em vacuum} state. It is not known with what probability negative energy densities occur in any other state, or for test functions other than a Gaussian, but it would be of interest to extend the analysis further. Investigations of similar 
phenomena in four dimensions can be found in~\cite{FewForRom:2012}. 
An application to two-dimensional dilaton quantum gravity was made in~\cite{CarlipMosnaPitelli:2011}, to argue that the 
positive energy tail causes strong focussing of light cones near the Planck scale.

\section{Other directions}\label{sect:other_dirns}

\subsection{Nonminimal coupling}

As mentioned above, the classical minimally coupled scalar field obeys the WEC by virtue
of a decomposition of the energy density as a sum of squares. This is not true for the 
nonminimally coupled field, and indeed the energy density can be made arbitrarily negative
at any given point (see, e.g., \cite{FordRoman_GSL:2001} for a discussion).  
It turns out that this behaviour is, nonetheless, constrained by locally averaged energy
conditions reminiscent of the QEIs~\cite{FewsterOsterbrink:2006}. For example, if
$\gamma$ is a complete causal geodesic with affine parameter $\lambda$ in a spacetime $\Mb$,
and the coupling constant is $\xi\in [0,1/4]$ then there is a bound
\[
\int_{\gamma} \mathrm{d} \lambda \ T_{ab} \dot{\gamma}^a \dot{\gamma}^b |g|^2
\geq -2\xi\int_{\gamma} \mathrm{d} \lambda \ \left\{|\partial_{\lambda} g|^{2}+\frac{1}{2} R_{ab}\dot{\gamma}^a \dot{\gamma}^b |g|^2 
-\left(\frac{1}{4}-\xi\right)R \dot{\gamma}^2 |g|^{2}\right\}\phi^{2} 
\]
for any solution $\phi$ to the nonminimally coupled Klein--Gordon equation, with corresponding
stress-energy tensor $T_{ab}$ and any $g\in\CoinX{\RR}$. In particular this result includes the case of conformal coupling. Note that the bound involves the field, but not its derivatives, while the 
quantity to be bounded involves field and derivatives, including some of second order. This
inequality therefore exhibits the `gain in derivatives' phenomenon that occurs in the 
G{\aa}rding inequalities of pseudodifferential operator theory. 

The corresponding quantum theory was discussed in~\cite{FewsterOsterbrink2008}
for the case of Minkowski space. It was found that the nonminimally coupled field can sustain large negative energy densities for long periods of time. The argument is the following: the failure of classical WEC 
allows the existence of one-particle states with negative energy density
near the origin, say. By a scaling
argument these can be taken to have any desired spacetime extent, although the magnitude is
correspondingly reduced. However, we may tensor together as many of these one-particle
states as we wish, with respect to which the energy density is additive (as this is a free theory).
Thus states of arbitrarily negative energy density can be sustained over arbitrarily large
spacetime volumes. 

However, there is a cost. The overall energy of these states is positive, and grows more rapidly
than the scales characterising the negative energy density effect produced. So the production
of negative energy density is inefficient in this sense. By modifying the QEI  arguments
discussed in these notes, one can establish QEI bounds for the nonminimally coupled field that are state-dependent~\cite{FewsterOsterbrink2008}. In these bounds the averaged energy density
is bounded below by state-independent terms together with terms that involve averages
of the expectation value of the Wick square of the field in the state of interest -- again demonstrating a `gain in derivatives' phenomenon. 
If one estimates these terms using the Hamiltonian operator, it again emerges that the
production of sustained negative energy densities only occurs when disproportionately
large positive energies are available. This is still a comparatively new development; further work, it is hoped, will clarify these issues.

\subsection{Interacting fields}

There is now a fairly complete theory of quantum energy inequalities
for free fields in globally hyperbolic spacetimes of any dimension
(although optimal bounds are lacking in general).
As we have seen, similar results hold in a large class of conformal
field theories in two dimensions. The situation for interacting fields
is more complicated, of course, and not so much is known. The
following remarks summarise the state of knowledge:
\begin{itemize}
\item One cannot expect state-independent QEIs to hold;
as mentioned, these can even fail in the nonminimally coupled
theory. Moreover, on physical grounds, we can expect that
long-lasting negative energy densities can be sustained by quantum fields as shown by the example of the Casimir effect, modelling the plates as certain states of a full interacting theory. A computation along
these lines was undertaken by Olum and Graham~\cite{OlumGraham03}; although there is a net
positive energy density near the `plates', their set-up maintains
a negative energy density near the mid-point between them.
However, it is possible that modified QEIs hold -- see~\cite[p.~176]{EvRom_book}
for some discussion.

\item In terms of positive results, the averaged null energy condition is known to hold 
in general two-dimensional quantum field theories~\cite{Verch_ANEC:2000}. In spacetime dimension
of $3$ or more, Bostelmann \& Fewster~\cite{BostelmannFewster:2009} have proved that for a wide class of theories obeying the `microscopic phase space condition' there are generally
state-dependent QI type results on quantities that are
`classically positive', i.e., arise as the leading term in the 
OPE of a sum of squares.
\end{itemize}

\subsection{Singularity theorems}

I originally motivated the energy conditions by reference to the
singularity theorems, in which they guarantee certain focussing
behaviour. An important question is whether or not the
QEI results provide sufficient control to guarantee that quantised
matter also obeys singularity theorems. 

Although this question is far from resolved, Fewster \& Galloway~\cite{FewsterGalloway:2011} have
recently shown that the hypotheses of the singularity theorems
can be weakened to accommodate bounds motivated by the 
QEIs. Unfortunately there is still a bit of a gap, because Hawking-style
singularity theorems, concerning congruences of timelike geodesics,
require the SEC (for which there is not a state-independent QEI) and Penrose-type results involve null geodesic
congruences (which are not suitable for QEI bounds). Nonetheless,
this is encouraging, and one can hope for more progress. 
For previous results along these lines see references in~\cite{FewsterGalloway:2011}.

\end{document}